\documentclass[12pt,eqsecnum]{article}
\usepackage[dvips]{graphicx}
\usepackage{amssymb}
\usepackage{amsmath}
\usepackage{ulem}
\usepackage[dvipsnames]{xcolor}
\usepackage{soul}
\usepackage{ascmac}
\usepackage{cite}
\usepackage{url}

\makeatletter

\@addtoreset{equation}{section}
\makeatletter

\newcommand{\qed}{\hbox{\rule[-2pt]{6pt}{6pt}}}
\newcommand{\D}{{\rm d}}

\newtheorem{Prop}{Proposition}

\newtheorem{lm}{Lemma}

\newcommand{\dalm}{\kern1pt\vbox{\hrule height 0.9pt\hbox{\vrule width
0.9pt\hskip 2.5pt\vbox{\vskip 5.5pt}\hskip 3pt\vrule width 0.3pt}\hrule height
0.3pt}\kern1pt}

\begin{document}

\begin{titlepage}
\vfill
\begin{flushright}
\today
\end{flushright}

\vfill
\begin{center}
\baselineskip=16pt
{\Large\bf
Conformally Schwarzschild cosmological black holes}
\vskip 0.0cm
{\large {\sl }}
\vskip 10.mm
{\bf Takuma Sato${}^{a}$, Hideki Maeda$^{b}$, and Tomohiro Harada${}^{a}$} \\

\vskip 1cm
{
${}^a$ Department of Physics, Rikkyo University, Toshima,
Tokyo 171-8501, Japan. \\
${}^b$ Department of Electronics and Information Engineering, Hokkai-Gakuen University, Sapporo 062-8605, Japan.\\
\texttt{stakuma@rikkyo.ac.jp, h-maeda@hgu.jp, harada@rikkyo.ac.jp}

}
\vspace{6pt}
\end{center}
\vskip 0.2in
\par
\begin{center}
{\bf Abstract}
\end{center}
\begin{quote}
We thoroughly investigate conformally Schwarzschild spacetimes in different coordinate systems to seek for physically reasonable models of a cosmological black hole.
We assume that a conformal factor depends only on the time coordinate and that the spacetime is asymptotically flat Friedmann-Lema\^{\i}tre-Robertson-Walker universe filled by a perfect fluid obeying a linear equation state $p=w\rho$ with $w>-1/3$.
In this class of spacetimes, the McClure-Dyer spacetime, constructed in terms of the isotropic coordinates, and the Thakurta spacetime, constructed in terms of the standard Schwarzschild coordinates, are identical and do not describe a cosmological black hole.
In contrast, the Sultana-Dyer and Culetu classes of spacetimes, constructed in terms of the Kerr-Schild and Painlev\'{e}-Gullstrand coordinates, respectively, describe a cosmological black hole.
In the Sultana-Dyer case, the corresponding matter field in general relativity can be interpreted as a combination of a homogeneous perfect fluid and an inhomogeneous null fluid, which is valid everywhere in the spacetime unlike Sultana and Dyer's interpretation.
In the Culetu case, the matter field can be interpreted as a combination of a homogeneous perfect fluid and an inhomogeneous anisotropic fluid.
However, in both cases, the total energy-momentum tensor violates all the standard energy conditions at a finite value of the radial coordinate in late times.
As a consequence, the Sultana-Dyer and Culetu black holes for $-1/3<w\le 1$ cannot describe the evolution of a primordial black hole after its horizon entry.
\vskip 1.mm
\noindent
{\bf Note added:} The present arXiv version corrects the published version (2022 Class. Quantum Grav. {\bf 39}, 215011) according to the corrigendum (2023 Class. Quantum Grav. {\bf 40}, 079501).
\vfill
\vskip 2.mm
\end{quote}
\end{titlepage}




\tableofcontents

\newpage

\section{Introduction}

A sufficiently isolated black hole in the universe should be well approximated by an asymptotically flat and stationary black-hole solution.
By the black-hole uniqueness theorem, it is known that the only asymptotically flat and stationary black hole with a regular and simply-connected event horizon in the Einstein-Maxwell system is the Kerr-Newman black hole.
(See~\cite{Chrusciel:2012jk} for a review.)
The discoveries of the black-hole thermodynamics~\cite{Bardeen:1973gs} and the Hawking radiation~\cite{Hawking:1975vcx} are major milestones in gravitational physics based on the uniqueness theorem, which means that one can understand quite general properties of an asymptotically flat and stationary black hole only through the study of the Kerr-Newman black hole.

However, the assumption of stationarity is not justified for dynamic black holes growing rapidly by absorbing the surrounding matter.
Moreover, when the radii of the event horizon and the cosmological horizon are relatively close, as in the case of primordial black holes just formed in the early universe, the assumption of asymptotic flatness is not justified either.
Such a {\it cosmological black hole} must be modeled by a dynamical black-hole solution which is asymptotic to an expanding universe. 
(See~\cite{Faraoni:2013aba,Faraoni:2018xwo} for reviews.)
The Schwarzschild-de~Sitter solution is surely the best known model of such a cosmological black hole, which is asymptotic to the de~Sitter universe.
Another example is the Einstein-Straus model~\cite{Einstein:1945id,Einstein:1946zz} (often discussed in the context of the ``Swiss-cheese'' model), which connects the Friedmann-Lema\^{\i}tre-Robertson-Walker (FLRW) solution with a dust fluid at a timelike hypersurface to the interior Schwarzschild spacetime.
Meanwhile, McVittie's spherically symmetric and asymptotically FLRW perfect-fluid solution~\cite{McVittie:1933zz} had been a candidate to describe a cosmological black hole after Nolan's study in his series of papers~\cite{Nolan:1998xs,Nolan:1999kk,Nolan:1999wf}.
Finally in 2012, Kaloper et al. showed that the maximally extended McVittie spacetime describes a cosmological black hole in the case where the scale factor obeys $a(t)\propto \exp(H_0 t)$ as $t\to \infty$ with a positive constant $H_0$~\cite{Kaloper:2010ec}. 
(See also~\cite{Lake:2011ni}.)

In addition to these solutions, several conformally Schwarzschild spacetimes have been proposed as cosmological black-hole models.
Thakurta proposed already in 1981 a conformally Schwarzschild spacetime in terms of the standard Schwarzschild coordinates with a conformal factor that depends only on the coordinate $t$~\cite{Thakurta1983}.
However, it has been exposed that this Thakurta spacetime has a curvature singularity at $r=2M$ and does not describe a cosmological black hole~\cite{Mello:2016irl,Harada:2021xze}.
In 2005, Sultana and Dyer successfully constructed a black-hole spacetime in terms of the Kerr-Schild coordinates which is asymptotic to the flat FLRW universe filled with a dust fluid~\cite{Sultana:2005tp}.
They showed that the corresponding energy-momentum tensor can be interpreted as a combination of a dust fluid and a null dust.
But unfortunately, their interpretation of matter is not valid in the whole spacetime because there is a region where tangent vectors of the orbits of a dust fluid and a null dust become complex.
Subsequently, McClure and Dyer investigated a conformally Schwarzschild spacetime constructed with the isotropic coordinates~\cite{McClure:2006kg}.
However, this spacetime can be transformed into the Thakurta spacetime by a coordinate transformation as pointed out in~\cite{Hammad:2018hhv}, and therefore it does not describe a cosmological black hole either.
In 2012, Culetu studied a conformally Schwarzschild spacetime constructed with the Painlev\'{e}-Gullstrand coordinates and found that it can be a model of a cosmological black hole and the corresponding matter field is an anisotropic fluid~\cite{Culetu:2012ih}.

A cosmological black-hole solution with physically reasonable matter fields would be a 
useful model in the study of primordial black holes, which are important candidates for dark matter.
The effect of the cosmic expansion may change the properties of black hole from the stationary case, and consequently affect the analysis for primordial black holes in cosmology.
In fact, the black-hole thermodynamics and the Hawking radiation of a cosmological black hole are highly non-trivial problems.
Conformally Schwarzschild cosmological black holes are quite useful in this context because the event horizon can be a conformal Killing horizon as well.
Jacobson and Kang defined the temperature $T_{\rm JKSD}$ of a conformal Killing horizon based on the argument that the temperature of a black hole should be conformally invariant because the Hawking radiations of a conformally coupled scalar field are identical from a Schwarzschild black hole and its conformal cousins sharing the same event horizon~\cite{Jacobson:1993pf}.
Sultana and Dyer also arrived the same definition of temperature in a different approach~\cite{SultanaDyer2004}.
The effect of the cosmic expansion on Hawking radiation has been analyzed using the Einstein-Strauss model as a model without matter accretion and the Sultana-Dyer black hole as a model with accretion~\cite{Saida:2007ru}.
Contrary to claims in~\cite{Jacobson:1993pf,SultanaDyer2004}, the effective temperature of the Sultana-Dyer black hole evaluated from the Hawking radiation is time-dependent and modified from $T_{\rm JKSD}$~\cite{Saida:2007ru}.
As these examples show, to identify cosmological black-hole solutions with a physically reasonable matter field and clarify their properties not only contributes to the fundamentals of black-hole physics but also to modern cosmology.

In the present paper, we will thoroughly investigate conformally Schwarzschild spacetimes with a conformal factor as a function only of the ``time'' coordinate in different coordinate systems of the Schwarzschild spacetime.
In particular, we will focus on the spacetimes which are asymptotic to the flat FLRW universe filled by a perfect fluid obeying a linear equation state $p=w\rho$ with $w>-1/3$.
The organization of the present paper is as follows.
In Sec.~\ref{sec:Preliminaries}, we will summarize mathematical results to study the conformally Schwarzschild spacetimes in the subsequent sections.
In Secs.~\ref{sec:SD} and~\ref{sec:Culetu}, we will clarify the global structures and the corresponding matter fields of the Sultana-Dyer class and the Culetu class of cosmological black holes, respectively.
Summary and discussions will be given in the final section.
In Appendix~\ref{app:others}, we present several non-conformally Schwarzschild spacetimes as other candidates of a more general cosmological black-hole spacetime.

Our conventions for curvature tensors are $[\nabla _\rho ,\nabla_\sigma]V^\mu ={{\cal R}^\mu }_{\nu\rho\sigma}V^\nu$ and ${\cal R}_{\mu \nu }={{\cal R}^\rho }_{\mu \rho \nu }$.
The signature of the Minkowski spacetime is $(-,+,+,+)$, and Greek indices run over all spacetime indices.
Throughout this paper, a dot on the scale factor $a$ denotes differentiation with respect to its argument.
We adopt units such that $c=G=\hbar=k_{\rm B}=1$.

\section{Preliminaries}
\label{sec:Preliminaries}

In this section, we summarize mathematical results to study spherically symmetric and conformally Schwarzschild spacetimes in the subsequent sections.

\subsection{Spherically symmetric spacetime}

The most general four-dimensional spherically symmetric spacetime $(M^4,g_{\mu\nu})$ is given by
\begin{align}
\D s^2 =g_{AB}(y)\D y^A\D y^B +R^2(y)\D \Omega^2,
\label{eq:ansatz}
\end{align}
where $y^A~(A=0,1)$ are coordinates in a two-dimensional Lorentzian spacetime $(M^2, g_{AB})$ and $\D \Omega^2:=\D \theta^2+\sin^2\theta \D \phi^2$.
The {\it areal radius} $R(y)(\ge 0)$ is a scalar on $(M^2, g_{AB})$.
The Einstein tensor of the spacetime with the metric (\ref{eq:ansatz}) is given by
\begin{align}
G_{\mu \nu }\D x^\mu \D x^\nu =&G_{AB}(y)\D y^A\D y^B +{\cal G}(y)R(y)^2\D \Omega^2,
\end{align} 
where a two-tensor $G_{AB}$ and a scalar ${\cal G}$ on $(M^2, g_{AB})$ are given by
\begin{align}
G_{AB}&=-2R^{-1}D_AD_BR-g_{AB}\left\{-2R^{-1}D^2R+R^{-2}[1-(DR)^2]\right\}, \label{GAB}\\
{\cal G}&=-\frac12{}^{(2)}{\cal R}+R^{-1} D^2R.
\end{align}
Here ${}^{(2)}{\cal R}$ is the Ricci scalar of $(M^2, g_{AB})$ and we have defined $(DR)^2:=g^{AB}(D_A R)(D_B R)$ and $D^2R:=g^{AB}D_AD_B R$ in terms of the covariant derivative $D_A$ on $(M^2, g_{AB})$.
Hence, in general relativity, the corresponding energy-momentum tensor $T_{\mu\nu}(=G_{\mu\nu}/8\pi)$ is given by 
\begin{align}
T_{\mu \nu }\D x^\mu \D x^\nu =&T_{AB}(y)\D y^A\D y^B +p_{\rm t}(y)R(y)^2\D \Omega^2 \label{Tmunu}
\end{align} 
with $T_{AB}=G_{AB}/8\pi$ and $p_{\rm t}={\cal G}/8\pi$.

For a spherically symmetric spacetime (\ref{eq:ansatz}), the Misner-Sharp quasi-local mass $m_{\rm MS}$~\cite{Misner:1964je} is defined by
\begin{align}
m_{\rm MS} := \frac{1}{2}R\left\{1-(DR)^2\right\},\label{MS-mass}
\end{align}
which satisfies
\begin{align}
D_A m_{\rm MS} =&4\pi R^{2}\left\{{T^B}_{A}(D_BR)- {T^B}_{B}(D_AR)\right\}.\label{dm}
\end{align} 
Properties of $m_{\rm MS}$ have been fully investigated in~\cite{Hayward:1994bu}.
The Misner-Sharp mass converges to the ADM mass at spacelike infinity in an asymptotically flat spacetime.

\subsubsection{Trapping horizon}
\label{sec:trapping}

In the present paper, we adopt spherical slicings to identify trapped round spheres and trapping horizons defined by Hayward~\cite{Hayward:1993wb}\footnote{We note that the term ``surface'' in Ref.~\cite{Hayward:1993wb} does not refer to a general closed two-surface but to a two-round sphere with constant $y^{A}~(A=0,1)$.}.
Let $k^\mu(\partial/\partial x^\mu)=k^A(\partial/\partial y^A)$ and $l^\mu(\partial/\partial x^\mu)=l^A(\partial/\partial y^A)$ be
two independent future-directed radial null vectors in the spacetime (\ref{eq:ansatz}) satisfying $k_\mu k^\mu=l_\mu l^\mu=0$ and $k_\mu l^\mu=-1$.
The expansions along those null vectors are given by
\begin{align}
\theta_+:=&\nabla_\mu k^\mu+l^\mu k^\nu\nabla_\nu k_\mu={\cal A}^{-1}{\cal L}_+{\cal A}=2R^{-1}k^AD_AR,\label{exp+}\\
\theta_-:=&\nabla_\mu l^\mu+k^\mu l^\nu\nabla_\nu l_\mu={\cal A}^{-1}{\cal L}_-{\cal A}=2R^{-1}l^AD_AR, \label{exp-}
\end{align}
where ${\cal A}:=4\pi R^2$ is the surface area of a two-round sphere with the areal radius $R$ given by $y^{A}={\rm constant}~(A=0,1)$ and we have defined ${\cal L}_+:=k^AD_A$ and ${\cal L}_-:=l^AD_A$.

In terms of $\theta_\pm$, a {\it trapped (untrapped) round sphere} is defined by a two-round sphere with $\theta_{+}\theta_{-}>(<)0$ and a {\it trapped (untrapped) region} is the union of all trapped (untrapped) round spheres.
A {\it marginal round sphere} is a two-round sphere with $\theta_{+}\theta_{-}=0$\footnote{Without spherical slicings, a (un)trapped {\it surface} and marginal {\it surface} are defined as a closed two-surface in a similar manner in terms of $\theta_\pm$ along two independent future-directed null vectors which are not necessarily radial.}.
Since the metric $g_{AB}$ on $(M^2, g_{AB})$ can be decomposed in terms of $k^A$ and $l^A$ as $g_{AB}=-k_A l_B-l_A k_B$, we obtain $\theta_+\theta_-=-2R^{-2}(DR)^2$.
Thus, an untrapped (trapped) region is given by $(D R)^2>(<)0$, or equivalently $R>(<)2m_{\rm MS}$ by Eq.~(\ref{MS-mass}).
A marginal round sphere is given by $(D R)^2=0$, or equivalently $R=2m_{\rm MS}$.

It is noted that a (un)trapped round sphere and marginal round sphere are defined with respect to a given SO(3) group defining the spherical symmetry as emphasized in Ref.~\cite{Bengtsson:2010tj}.
Such a SO(3) group is unique in generic spherically symmetric spacetimes, however, there are spacetimes where the SO(3) group is not unique, such as flat, (anti-)de~Sitter and Friedmann-Lema\^{i}tre-Robertson-Walker spacetimes.
In such a spacetime with higher symmetry, the locations of trapping horizons and (un)trapped regions depend on the choice of the SO(3) group.

Using the degrees of freedom to interchange such that $\theta_+\leftrightarrow \theta_-$, one may set $\theta_{+}=0$ on a marginal round sphere without loss of generality.
Then, a marginal round sphere is said to be {\it future} if $\theta_-<0$, {\it past} if
$\theta_{-}>0$, {\it bifurcating} if $\theta_-=0$, {\it outer} if
${\cal L}_-\theta_{+}<0$, {\it inner} if ${\cal L}_-\theta_{+}>0$ and {\it
degenerate} if ${\cal L}_-\theta_{+}=0$.
Finally, a {\it trapping horizon} is the closure of a hypersurface foliated by marginal round spheres~\cite{Hayward:1993wb}\footnote{The original definition by Hayward~\cite{Hayward:1993wb,Hayward:1997jp} requires a foliation only by future or past and outer or inner marginal round spheres.}.
All the possible types of trapping horizon are summarized in Table~\ref{table:TH-class}.
\begin{table}[htb]
\begin{center}
\caption{\label{table:TH-class} Types of trapping horizon $\Sigma$ given by $\theta_+=0$ and their interpretations.}
\scalebox{0.85}{
\begin{tabular}{|c|c|c|c|c|}
\hline
& ${\cal L}_{-}\theta_{+}|_\Sigma>0$ & ${\cal L}_{-}\theta_{+}|_\Sigma=0$ & ${\cal L}_{-}\theta_{+}|_\Sigma<0$ \\ \hline\hline
$\theta_-|_\Sigma>0$ & Past inner (Cosmological) & Past degenerate & Past outer (White hole) \\ \hline
$\theta_-|_\Sigma=0$ & Bifurcating inner & Bifurcating degenerate & Bifurcating outer \\ \hline
$\theta_-|_\Sigma<0$ & Future inner (Anti-cosmological) & Future degenerate & Future outer (Black hole) \\
\hline
\end{tabular}
}
\end{center}
\end{table}

The inequality $\theta_-|_\Sigma<0$ means that the ingoing null rays converge on the trapping horizon $\Sigma$, while the inequality ${\cal L}_-\theta_{+}|_\Sigma<0$ means that the outgoing null rays are instantaneously parallel on $\Sigma$ but diverging just outside $\Sigma$ and converging just inside.
Thus, a future outer trapping horizon corresponds to a black-hole horizon among others~\cite{Hayward:1993wb,Hayward:1997jp}.
On the other hand, a past inner trapping horizon and a past outer trapping horizon correspond to a cosmological horizon and a white-hole horizon, respectively.
The inner Cauchy horizon in the non-extreme Reissner-Nordstr\"om black hole is a future inner trapping horizon.
\begin{figure}[htbp]
\begin{center}
\includegraphics[width=0.9\linewidth]{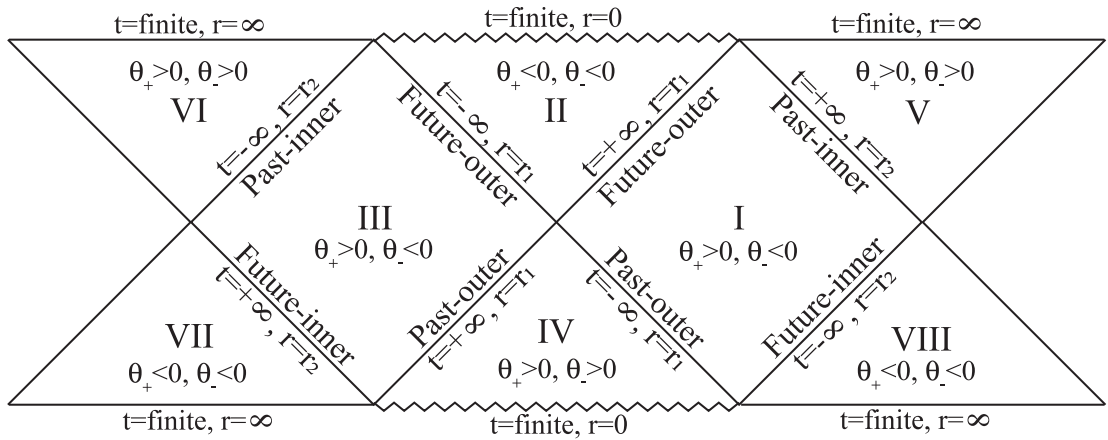}
\caption{\label{Schwarzschild-dS} A Penrose diagram of the maximally extended Schwarzschild-de~Sitter black hole (\ref{Sc-dS}).}
\end{center}
\end{figure}

As an example, Fig.~\ref{Schwarzschild-dS} exhibits four different types of trapping horizons in the Schwarzschild-de~Sitter spacetime with a metric given by
\begin{align}
\label{Sc-dS}
\begin{aligned}
&\D s^2=-f(r)\D t^2+f(r)^{-1}\D r^2+r^2\D \Omega^2, \\
&f(r)=1-\frac{2M}{r}-\frac13\Lambda r^2,
\end{aligned}
\end{align}
where $M$ and $\Lambda(>0)$ satisfy $M>1/(3\sqrt{\Lambda})$ so as to generate two Killing horizons at $r=r_1(>0)$ and $r_2(>r_1)$, which are trapping horizons as well.
Here it should be noted that all the results in Sec.~9 in the textbook~\cite{he1973} cannot be applied directly to the Schwarzschild-de~Sitter spacetime and other cosmological black-hole spacetimes.
For example, the fact that outer trapped round spheres in the regions VII and VIII in Fig.~\ref{Schwarzschild-dS} can be seen from the future null infinity does not contradict to Proposition~9.2.8 in~\cite{he1973}.
This is because these black-hole spacetimes do not satisfy the condition (4) in page 222, so that they are not weakly asymptotically simple and empty, which is an assumption in Sec.~9 in~\cite{he1973}.

Here we prove that the location of a trapping horizon and its type are invariant concepts with spherical slicings.
The following proposition is a generalization of the claim in~\cite{Faraoni:2016xgy} and essentially the same as Result 6.1 in~\cite{Bengtsson:2010tj}, whereas we provide an explicit proof on the independence from the choice of the pair of the future-directed radial null vector fields\footnote{Reference~\cite{Bengtsson:2010tj} came within the attention of the authors after the publication of the present paper.}.

\begin{Prop}
\label{Prop:signature}
With spherical slicings, the location of a trapping horizon $\Sigma$ and its type with respect to a {\it given} SO(3) group of isometry in a spherically symmetric spacetime (\ref{eq:ansatz}) are invariant under not only coordinate transformations on $(M^2, g_{AB})$ but also the freedom in choosing a pair of future-directed radial null vectors ${\boldsymbol k}=k^A\partial/\partial y^A$ and ${\boldsymbol l}=l^A\partial/\partial y^A$, which are regular on $\Sigma$.
\end{Prop}
{\it Proof:}
Since the null expansions $\theta_{\pm}$ and their derivatives ${\cal L}_{\pm }\theta_{\mp }$ are scalars on $(M^{2}, g_{AB})$ by definitions (\ref{exp+}) and (\ref{exp-}), the values of $\theta_{\pm}$ and ${\cal L}_{\pm }\theta_{\mp }$ at each point $p\in (M^{4}, g_{\mu\nu})$ are unchanged under coordinate transformations on $(M^{2}, g_{AB})$ if we fix a pair $\{{\boldsymbol k},{\boldsymbol l}\}$. 
Therefore, the location of a trapping horizon and its type are invariant under coordinate transformations on $(M^{2}, g_{AB})$ for a given $\{{\boldsymbol k},{\boldsymbol l}\}$.
So, what is to prove is independence from the choice of the pair of future-directed radial null vectors.

Let ${\boldsymbol k}=k^A(\partial/\partial y^A)$ and ${\boldsymbol l}=l^A(\partial/\partial y^A)$ be the original choice, where $k^A$ and $l^A$ are assumed to be finite on trapping horizons.
Using the degrees of freedom to interchange such that ${\boldsymbol k}\leftrightarrow {\boldsymbol l}$, one may set $\theta_{k}=0$ on a trapping horizon without loss of generality, where $\theta_{k}:= 2R^{-1}k^{A}D _{A}R$.
Now let ${\tilde{\boldsymbol k}}={\tilde k}^A(\partial/\partial y^A)$ and ${\tilde{\boldsymbol l}}={\tilde l}^A(\partial/\partial y^A)$ be a new choice of future-directed radial null vectors in the same coordinates on $(M^2,g_{AB})$. 
Since there are only two independent future-directed radial null vectors at each spacetime point up to a multiplication factor and we impose $\tilde{k}^{A}\tilde{l}_{A}=k^{A}l_{A}=-1$, $\{{\tilde k}^A,{\tilde l}^A\}$ may be represented as $\tilde{k}^{A}=\beta k^{A}$ and $\tilde{l}^{A}=\beta^{-1}l^{A}$, where $\beta=\beta(y)$ is a scalar on $(M^{2}, g_{AB})$ and assumed to be positive for $\{{\tilde{\boldsymbol k}},{\tilde{\boldsymbol l}}\}$ to be future directed.
Then, we obtain
\begin{align}
&{\theta}_{{\tilde k}}=\beta \theta_{k},\qquad {\theta}_{{\tilde l}}=\beta^{-1}\theta_{l},\label{eq:thetatildepm}\\
&{\cal L}_{\tilde{l}}\theta_{\tilde{k}}={\cal L}_{l}\theta_{k}+\beta^{-1}(l^{A}D_{A}\beta) \theta_{k}\label{eq:Lthetatildepm}
\end{align}
and assume that the positive function $\beta$ is $C^1$ on a trapping horizon $\Sigma$ defined by $\theta_{k}=0$.
Equation~(\ref{eq:thetatildepm}) shows that ${\theta}_{{\tilde k}}|_\Sigma=0$ is equivalent to $\theta_{k}|_\Sigma=0$ and the signs of ${\theta}_{{\tilde l}}$ and $\theta_{l}$ are the same on $\Sigma$.
In addition, Eq.~(\ref{eq:Lthetatildepm}) shows ${\cal L}_{\tilde{l}}\theta_{\tilde{k}}|_\Sigma={\cal L}_{l}\theta_{k}|_\Sigma$.
Hence, the location of a trapping horizon and its type are independent from the choice of the pair of future-directed radial null vectors.
\qed

It is noted that we need to fix the SO(3) group defining the spherical symmetry in Proposition~\ref{Prop:signature} as emphasized in Ref.~\cite{Bengtsson:2010tj}.
We also note that the regularity assumption on ${\boldsymbol k}$ and ${\boldsymbol l}$ on the trapping horizon $\Sigma$ is indispensable to prove Proposition~\ref{Prop:signature}.
As a concrete example, let us consider the Schwarzschild spacetime in the ingoing Eddington-Finkelstein coordinates
\begin{align}
\D s^2=-f(r)\D v^2+2\D v\D r+r^2\D \Omega^2 
\end{align}
with $f(r)=1-2M/r$, of which trapping horizon $\Sigma$ determined by $f(r)=0$ coincides with the event horizon.
Consider a pair of future-directed radial null vectors
\begin{align}
{\boldsymbol k}=&\frac{1}{\sqrt{2}}\biggl(2\frac{\partial}{\partial v}+f\frac{\partial}{\partial r}\biggl),\qquad {\boldsymbol l}=-\frac{1}{\sqrt{2}}\frac{\partial}{\partial r},
\end{align}
of which components are finite on $\Sigma$ and satisfy $k_\mu k^\mu=0=l_\mu l^\mu$ and $k_\mu l^\mu=-1$.
The expansions $\theta_k=\sqrt{2}f/r$ and $\theta_l=-\sqrt{2}/r$ show $\theta_l|_\Sigma<0$, so that the trapping horizon is of the future type.
In contrast, for a different pair ${\tilde{\boldsymbol k}}=f^{-1/2}{\boldsymbol k}$ and ${\tilde{\boldsymbol l}}=f^{1/2}{\boldsymbol l}$, of which components are
\begin{align}
{\tilde{\boldsymbol k}}=&\frac{1}{\sqrt{2}}\biggl(\frac{2}{f^{1/2}}\frac{\partial}{\partial v}+f^{1/2}\frac{\partial}{\partial r}\biggl),\qquad {\tilde{\boldsymbol l}}=-\frac{f^{1/2}}{\sqrt{2}}\frac{\partial}{\partial r},
\end{align}
the expansions $\theta_{\tilde k}=\sqrt{2}f^{1/2}/r$ and $\theta_{\tilde l}=-\sqrt{2}f^{1/2}/r$ provide a wrong answer $\theta_{\tilde l}|_\Sigma=0$.
This is a consequence of $\tilde{\boldsymbol{k}}$ being singular on $\Sigma$.

Lastly, the contrapositions of the following propositions~\cite{Hayward:1993wb} are useful to identify the region where the null energy condition or the dominant energy condition is violated. (See also~\cite{Nozawa:2007vq,Maeda:2007uu}.)
\begin{Prop}
\label{prop:sig}
Under the null energy condition, an outer (inner) trapping horizon in the spacetime (\ref{eq:ansatz}) is non-timelike (non-spacelike).
\end{Prop}
\begin{Prop}
\label{arealaw}
Under the null energy condition, the area of a future outer (inner) trapping horizon in the spacetime (\ref{eq:ansatz}) is non-decreasing (non-increasing) along its generator.
\end{Prop}
\begin{Prop}
\label{th:monotonicity}
Under the dominant energy condition, the Misner-Sharp mass $m_{\rm MS}$ is non-decreasing (non-increasing) in any outgoing (ingoing) spacelike or null direction on an untrapped round sphere.
\end{Prop}

\subsubsection{Kodama vector and Misner-Sharp mass}

The general spherically symmetric spacetime (\ref{eq:ansatz}) admits the Kodama vector $K^\mu(\partial/\partial x^\mu)=K^A(\partial/\partial y^A)$~\cite{Kodama:1979vn}.
Here a vector $K^A$ on $(M^2, g_{AB})$ is defined by 
\begin{align}
K^A :=-\epsilon ^{AB}D_B R~~\biggl(\Leftrightarrow~~D_BR=-\epsilon_{BA}K^A\biggl), \label{kodamavector}
\end{align}
where $\epsilon_{AB}$ is a volume two-form on $(M^2, g_{AB})$ satisfying
\begin{align}
\begin{aligned}
&\epsilon_{01}(=-\epsilon_{10})=\sqrt{-\det(g_{AB})},\\
&\epsilon_{AB}\epsilon^{CD}=-(\delta^C_A\delta^D_{B}-\delta^C_B\delta^D_{A})
\end{aligned}
\end{align}
and hence $\epsilon^{01}=-1/\epsilon_{01}$.
$K^A$ is orthogonal to $D_AR$ and the minus sign in the definition of $K^A$ in Eq.~(\ref{kodamavector}) is to make $K^A$ future-pointing.
The expression $K^\mu K_\mu =K^A K_A =-(D R)^2$ shows that $K^\mu$ is timelike and spacelike in untrapped regions and trapped regions, respectively, and it is null on trapping horizons.
Since the Kodama vector $K^\mu$ reduces to a hypersurface-orthogonal timelike Killing vector if the spacetime is static, it generates a preferred time direction in untrapped regions in the general spherically symmetric spacetime.

In fact, the Misner-Sharp mass (\ref{MS-mass}) is a locally conserved charge along an energy current vector $J^\mu:=-{T^\mu}_\nu K^\nu$ associated with a Kodama observer (of which orbit is timelike only in untrapped regions).
Here we have $J^\mu(\partial/\partial x^\mu)=J^A(\partial/\partial y^A)$, where
\begin{align}
J^A =-{T^A}_B K^B\biggl(= -\frac{1}{8\pi}G^{AB}K_{B}\biggl).\label{kodamacurrent}
\end{align}
One can show that $J^\mu$ is divergence-free ($\nabla_\mu J^\mu=0$)~\cite{Maeda:2007uu} and then the integral of $-J^\mu$ over a spacelike hypersurface $\Pi$ with boundary gives an associated charge $Q_J := -\int _\Pi J^\mu \D \Pi _\mu$, where $\D \Pi _\mu$ is a directed surface element on $\Pi$\footnote{Note that the minus sign in the definition of $Q_J$ is due to the Minkowski signature $(-,+,+,+)$.}.
One may use another expression $\D \Pi _\mu=u_\mu \D \Pi$ with a future-directed unit normal $u^\mu$ to $\Pi$ and a surface element $\D \Pi$ on $\Pi$.
In fact, the charge $Q_J$ is identical to $m_{\rm MS}$ up to an integration constant.
Suppose that $\Pi$ is defined by $y^0=t_0=$constant and then we have $u_\mu\D x^\mu=-(1/\sqrt{-g^{00}})\D y^0$.
Then, using $g^{00}=g_{11}/\det(g_{AB})$, we obtain
\begin{align}
Q_J =&\int \partial_1 m_{\rm MS}(t_0,y^1)\D y^1=\left[m_{\rm MS}(t_0,y^1)\right]_{y^1=b_1}^{y^1=b_2},
\end{align}
where $y^1=b_1$ and $y^1=b_2$ correspond to boundaries on $\Pi$.
It is reasonable to set $y^1=b_1$ correspond to a regular center if it exists.

Here we note that the Kodama vector $K^\mu$ itself is divergence-free ($\nabla_\mu K^\mu=0$)~\cite{Maeda:2007uu}, so that it is also a locally conserved current.
Its associated charge $Q_K := -\int _\Pi K^\mu \D \Pi _\mu$ is the volume $4\pi R^3/3$ inside a two-round sphere with the areal radius $R$.

\subsubsection{Compatible matter field in general relativity}

In the present paper, we will consider a matter field compatible with a given spacetime $(M^4,g_{\mu\nu})$ in general relativity, of which energy-momentum tensor is determined through the Einstein equations as ${T}_{\mu\nu}=G_{\mu\nu}/(8\pi)$.
According to the Hawking-Ellis classification, an energy-momentum tensor $T_{\mu\nu}$ is classified into four types depending on the properties of its eigenvectors as shown in Table~\ref{table:scalar+1}~\cite{he1973,Martin-Moruno:2017exc,Maeda:2018hqu}.
In general relativity, any spherically symmetric spacetime is compatible with an energy-momentum tensor of type I, II, or IV.
\begin{table}[htb]
\begin{center}
\caption{\label{table:scalar+1} Eigenvectors of type-I--IV energy-momentum tensors.}
\scalebox{0.90}{
\begin{tabular}{|c|c|c|c|}
\hline
Type & Eigenvectors \\ \hline\hline
I & 1 timelike, 3 spacelike \\ \hline
II & 1 null (doubly degenerated), $2$ spacelike \\ \hline
III & 1 null (triply degenerated), 1 spacelike \\ \hline
IV & 2 complex, $2$ spacelike \\
\hline
\end{tabular}
}\end{center}
\end{table}

Let $\{{E}_\mu^{(a)}\}~(a=0,1,2,3)$ be orthonormal basis one-forms in the local Lorentz frame satisfying
\begin{equation}
{E}^\mu_{(a)}{E}_{(b)\mu}=\eta_{(a)(b)}=\mbox{diag}(-1,1,1,1).
\end{equation}
Here $\eta_{(a)(b)}$ is the metric in the local Lorentz frame and the spacetime metric $g_{\mu\nu}$ is given by $g_{\mu\nu}=\eta_{(a)(b)}E^{(a)}_{\mu}E^{(b)}_{\nu}$.
$\eta_{(a)(b)}$ and its inverse $\eta^{(a)(b)}$ are respectively used to lower and raise the indices with brackets.

In the most general spherically symmetric spacetime (\ref{eq:ansatz}), one may introduce $\{{E}_\mu^{(a)}\}$ such that
\begin{align}
{E}_\mu^{(\alpha)}\D x^\mu={E}_A^{(\alpha)}\D y^A,\qquad {E}_\mu^{(2)}\D x^\mu=R\D\theta,\qquad {E}_\mu^{(3)}\D x^\mu=R\sin\theta\D\phi, \label{standard-basis}
\end{align}
where the basis one-forms $\{{E}_A^{(\alpha)}\}~(\alpha=0,1)$ on $(M^2, g_{AB})$ satisfy
\begin{equation}
{E}^A_{(\alpha)}{E}_{(\beta)A}=\eta_{(\alpha)(\beta)}=\mbox{diag}(-1,1).
\end{equation}
Here $\eta_{(\alpha)(\beta)}$ is the metric in the local Lorentz frame on $(M^2, g_{AB})$.
Then, non-zero components of $G^{(a)(b)}$ are $G^{(0)(0)}$, $G^{(0)(1)}(=G^{(1)(0)})$, $G^{(1)(1)}$, and $G^{(2)(2)}(=G^{(3)(3)})$.
In a region with $G^{(0)(1)}=0$, the Hawking-Ellis type of the corresponding energy-momentum tensor (\ref{Tmunu}) is of type I.
In a region with $G^{(0)(1)}\ne 0$, we can use the following lemma~\cite{Maeda:2022vld}.
\begin{lm}
\label{Prop:HE-type}The Hawking-Ellis type of the energy-momentum tensor (\ref{Tmunu}) is type I if $T^{(0)(1)}=0$.
If $T^{(0)(1)}\ne 0$, it is determined as
\begin{align}
&(T^{(0)(0)}+T^{(1)(1)})^2> 4(T^{(0)(1)})^2~~\Rightarrow~~\mbox{\rm Type~I},\\
&(T^{(0)(0)}+T^{(1)(1)})^2= 4(T^{(0)(1)})^2~~\Rightarrow~~\mbox{\rm Type~II},\\
&(T^{(0)(0)}+T^{(1)(1)})^2< 4(T^{(0)(1)})^2~~\Rightarrow~~\mbox{\rm Type~IV}.
\end{align}
\end{lm}

The standard energy conditions consist of the {\it null} energy condition (NEC), {\it weak} energy condition (WEC), {\it dominant} energy condition (DEC), and {\it strong} energy condition (SEC).
Using the local Lorentz transformation, one can write $T^{(a)(b)}$ in a canonical form~\cite{he1973,Martin-Moruno:2017exc,Maeda:2018hqu}.
In a spherically symmetric spacetime, the canonical form of type I is
\begin{equation}
\label{T-typeI}
T^{(a)(b)}=\mbox{diag}(\rho,p_{\rm r},p_{\rm t},p_{\rm t}).
\end{equation}
Here $\rho$, $p_{\rm r}$, and $p_{\rm t}$ are interpreted as the energy density, radial pressure, and tangential pressure, respectively, and equivalent expressions of the standard energy conditions are
\begin{align}
\mbox{NEC}:&~~\rho+p_{\rm r}\ge 0~~\mbox{and}~~\rho+p_{\rm t}\ge 0,\label{NEC-I}\\
\mbox{WEC}:&~~\rho\ge 0\mbox{~in addition to NEC},\label{WEC-I}\\
\mbox{DEC}:&~~\rho-p_{\rm r}\ge 0~~\mbox{and}~~\rho-p_{\rm t}\ge 0\mbox{~in addition to WEC},\label{DEC-I}\\
\mbox{SEC}:&~~\rho+p_{\rm r}+2p_{\rm t}\ge 0\mbox{~in addition to NEC}.\label{SEC-I}
\end{align}
The canonical form of type II is
\begin{eqnarray}
\label{T-typeII}
T^{(a)(b)}=\left(
\begin{array}{cccc}
\rho+\nu &\nu&0&0\\
\nu&-\rho+\nu&0&0\\
0&0&p_{\rm t}&0 \\
0&0&0 &p_{\rm t}
\end{array}
\right)
\end{eqnarray}
with $\nu\ne 0$ and equivalent expressions of the standard energy conditions are
\begin{align}
\mbox{NEC}:&~~\nu\ge 0\mbox{~and~}\rho+p_{\rm t}\ge 0,\label{NEC-II}\\
\mbox{WEC}:&~~\rho\ge 0\mbox{~in addition to NEC},\label{WEC-II}\\
\mbox{DEC}:&~~\rho-p_{\rm t}\ge 0\mbox{~in addition to WEC},\label{DEC-II}\\
\mbox{SEC}:&~~p_{\rm t}\ge 0\mbox{~in addition to NEC}.\label{SEC-II}
\end{align}
The type-IV energy-momentum tensor violates all the standard energy conditions.

For example, let us consider the flat FLRW spacetime with a conformal time $t$, in which the line element is written as:
\begin{align}
\D s^2=a(t)^2\left(-\D t^2+\D r^2+r^2\D \Omega^2\right).\label{flat-FLRW}
\end{align}
Adopting the following orthonormal basis one-forms:
\begin{align}
E_\mu^{(0)}\D x^\mu=-a\D t,\quad E_\mu^{(1)}\D x^\mu=a\D r,\quad E_\mu^{(2)}\D x^\mu=ar\D \theta, \quad E_\mu^{(3)}\D x^\mu=ar\sin\theta \D \phi,
\end{align}
we obtain
\begin{align}
G^{(0)(0)}=&3\frac{{\dot a}^2}{a^4},\qquad G^{(1)(1)}=G^{(2)(2)}=G^{(3)(3)}=-2\frac{{\ddot a}}{a^3}+\frac{{\dot a}^2}{a^4},
\end{align}
so that the corresponding energy-momentum tensor $T_{\mu\nu}$ is of the Hawking-Ellis type I (\ref{T-typeI}) with $p_{\rm r}=p_{\rm t}(\equiv p)$, where
\begin{align}
8\pi \rho=3\frac{{\dot a}^2}{a^4},\qquad 8\pi p=-2\frac{{\ddot a}}{a^3}+\frac{{\dot a}^2}{a^4}.
\end{align}
This matter field can be interpreted as a perfect fluid in the comoving coordinates:
\begin{align}
&{T}_{\mu\nu}=(\rho+p)u_\mu u_\nu+pg_{\mu\nu},\label{T-pf}\\
&u^\mu\frac{\partial}{\partial x^\mu}=\frac{1}{a}\frac{\partial}{\partial t}.
\end{align}
In the present paper, we assume that the scale factor obeys a power law $a(t)=a_0|t|^{\alpha}$, where $a_0$ and $\alpha$ are constants.
In general relativity, such a conformal factor with $\alpha=2/(3w+1)$ is a solution for a perfect fluid obeying an equation of state $p=w\rho$.
Then we have
\begin{align}
\begin{aligned}
&8\pi a^2(\rho+p)=\frac{2\alpha(\alpha+1)}{\tau^2},\qquad 8\pi a^2\rho=3\frac{\alpha^2}{\tau^2},\\
&8\pi a^2(\rho-p)=\frac{2\alpha(2\alpha-1)}{\tau^2},\qquad 8\pi a^2(\rho+3p)=\frac{6\alpha}{\tau^2}
\end{aligned}
\end{align}
and the standard energy conditions are summarized in Table~\ref{table:EC-FLRW}.
\begin{table}[htb]
\begin{center}
\caption{\label{table:EC-FLRW} Energy conditions for the flat FLRW spacetime (\ref{flat-FLRW}) with $a(t)=a_0|t|^{\alpha}$.}
\begin{tabular}{|c|c|c|c|c|c|}
\hline
& $\alpha\le -1$ & $-1<\alpha< 0$ & $\alpha=0$ & $0<\alpha < 1/2$ & $\alpha \ge 1/2$ \\ \hline\hline
NEC & \checkmark & $\times$ & \checkmark & \checkmark & \checkmark \\ \hline
WEC & \checkmark & $\times$ & \checkmark & \checkmark & \checkmark \\ \hline
DEC & \checkmark & $\times$ & \checkmark & $\times$ & \checkmark \\ \hline
SEC & $\times$ & $\times$ & \checkmark & \checkmark & \checkmark \\
\hline
\end{tabular}

\end{center}
\end{table}

We will also use the following proposition~\cite{Maeda:2022vld} in the subsequent sections.
\begin{Prop}
\label{Prop:EC-criteriaI}
For an energy-momentum tensor (\ref{Tmunu}) in an orthonormal frame, all the standard energy conditions are violated if $(T^{(0)(0)}+T^{(1)(1)})^2< 4(T^{(0)(1)})^2$ or $T^{(0)(0)}+T^{(1)(1)}<0$ is satisfied.
If $(T^{(0)(0)}+T^{(1)(1)})^2\ge 4(T^{(0)(1)})^2$ and $T^{(0)(0)}+T^{(1)(1)}\ge 0$ hold, equivalent expressions of the standard energy conditions are given by
\begin{align}
\mbox{NEC}:&~~T^{(0)(0)}-T^{(1)(1)}+2p_{\rm t}+\sqrt{(T^{(0)(0)}+T^{(1)(1)})^2-4(T^{(0)(1)})^2}\ge 0,\\
\mbox{WEC}:&~~T^{(0)(0)}-T^{(1)(1)}+\sqrt{(T^{(0)(0)}+T^{(1)(1)})^2-4(T^{(0)(1)})^2}\ge 0 \nonumber \\
&\mbox{~~in addition to NEC},\\
\mbox{DEC}:&~T^{(0)(0)}-T^{(1)(1)}-2p_{\rm t}+\sqrt{(T^{(0)(0)}+T^{(1)(1)})^2-4(T^{(0)(1)})^2}\ge 0 \nonumber \\
&~~\mbox{and}~~T^{(0)(0)}-T^{(1)(1)}\ge 0\mbox{~~in addition to WEC},\\
\mbox{SEC}:&~~2p_{\rm t}+\sqrt{(T^{(0)(0)}+T^{(1)(1)})^2-4(T^{(0)(1)})^2}\ge 0\mbox{~~in addition to NEC}.
\end{align}
\end{Prop}

\subsection{The Schwarzschild spacetime revisited}
\begin{figure}[htbp]
\begin{center}
\includegraphics[width=0.7\linewidth]{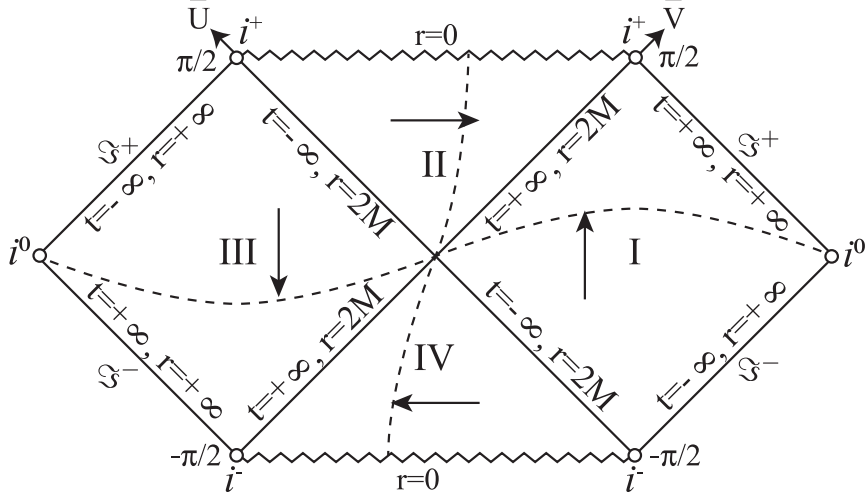}
\caption{\label{SingleBH} A Penrose diagram of the maximally extended Schwarzschild black hole. 
A dashed curve represents a constant $t$ hypersurface in the Schwarzschild coordinates (\ref{Sc1}) and an arrow shows the increasing direction of $t$.
${\bar U}$ and ${\bar V}$ are compactified Kruskal-Szekeres coordinates (\ref{barUV-def}).
$i^+ (i^-)$ is a future (past) timelike infinity and $i^0$ is a spacelike infinity. $\Im^{+} (\Im^{-})$ is a future (past) null infinity.} 
\end{center}
\end{figure}

The Schwarzschild vacuum solution is written in the most well-known diagonal coordinates $\{t,r,\theta,\phi\}$ as
\begin{eqnarray}
\D s^2=-\biggl(1-\frac{2M}{r}\biggl)\D t^2+\biggl(1-\frac{2M}{r}\biggl)^{-1}\D r^2+r^2\D \Omega^2, \label{Sc1}
\end{eqnarray}
where $M$ is a constant.
We refer to the coordinates (\ref{Sc1}) as the Schwarzschild coordinates.
The domains of coordinates $\{t,\theta,\phi\}$ are $-\infty<t<\infty$, $0\le\theta\le\pi$, and $0\le \phi<2\pi$.
Hereafter, we assume that $M$ is positive corresponding to the black-hole case and then $r=0$ is a spacelike curvature singularity.
In the coordinate system (\ref{Sc1}), the event horizon $r=2M$ is a coordinate singularity.
As a result, the domain of the coordinate $r$ is given by $0<r<2M$ and $2M<r<\infty$.
Therefore, although the Schwarzschild coordinates (\ref{Sc1}) cover the region I, II, III, or IV in the maximally extended Schwarzschild spacetime shown in Fig.~\ref{SingleBH}, they do not cover the event horizon $r=2M$.

\noindent
{\bf The Eddington-Finkelstein and the Kruskal-Szekeres coordinates}

In the spacetime with the metric (\ref{Sc1}), we introduce an outgoing null coordinate $u$ and an ingoing null coordinate $v$ such that
\begin{eqnarray}
u:=t-r_*, \qquad v:=t+r_*,\label{uv-def}
\end{eqnarray}
where $r_*$ is the tortoise coordinate defined by
\begin{eqnarray}
r_{\ast}:=r+2M\ln {\biggl|\frac{r}{2M}-1\biggl|}.\label{r*-def}
\end{eqnarray}
The tortoise coordinate satisfies $\D r_*=f(r)^{-1}\D r$ and 
\begin{align}
&\lim_{r\to 0}r_*=2M\ln|2M|, \qquad \lim_{r\to 2M}r_*\to -\infty,\qquad \lim_{r\to +\infty}r_*\to +\infty.
\end{align}
The domains of $u$ and $v$ are $-\infty<u<\infty$ and $-\infty<v<\infty$, respectively.
In the outgoing Eddington-Finkelstein coordinates $\{u,r\}$ and the ingoing Eddington-Finkelstein coordinates $\{v,r\}$ on $(M^2,g_{AB})$, the line element in the Schwarzschild spacetime is written as
\begin{align}
\D s^2=&-\biggl(1-\frac{2M}{r}\biggl)\D u^2-2\D u\D r+r^2\D \Omega^2, \label{Sc2}\\
\D s^2=&-\biggl(1-\frac{2M}{r}\biggl)\D v^2+2\D v\D r+r^2\D \Omega^2, \label{Sc3}
\end{align}
respectively.
Since the metric and its inverse are both regular at the event horizon $r=2M$, the domain of $r$ is $0<r<\infty$ in the Eddington-Finkelstein coordinates (\ref{Sc2}) and (\ref{Sc3}).
Nevertheless, they cover a half of the maximally extended Schwarzschild spacetime with $M>0$.
In fact, the coordinates (\ref{Sc2}) cover the region I$+$IV or II$+$III in Fig.~\ref{SingleBH}, while the coordinates (\ref{Sc3}) cover the region I$+$II or III$+$IV.

In order to cover the maximally extended Schwarzschild spacetime in a single coordinate system, we introduce the Kruskal-Szekeres coordinates $\{U,V\}$ on $(M^2,g_{AB})$ such that
\begin{eqnarray}
U = -e^{-u/(4M)},\qquad V = e^{v/(4M)}, \label{UV-def}
\end{eqnarray}
in which the line element in the Schwarzschild spacetime is written as
\begin{align}
\D s^2 = - \frac{32M^3}{r} e^{-r/(2M)} \D U \D V + r(U,V)^2 \D \Omega^2, \label{KS-coors}
\end{align}
where the function $r(U,V)$ is implicitly given from
\begin{align}
UV = \biggl(1 - \frac{r}{2M}\biggl) e^{r/(2M)}. \label{UV-r}
\end{align}
Since $U$ and $V$ defined by Eq.~(\ref{UV-def}) satisfy $U<0$ and $V>0$, the relation (\ref{UV-r}) shows that the coordinates $\{U,V\}$ cover the region $r>2M$.
Nevertheless, since the metric (\ref{KS-coors}) is analytic at $UV=0$ corresponding to $r=2M$, the spacetime is analytically extended into the region of $U\ge 0$ or $V\le 0$.
Accordingly, the Kruskal-Szekeres coordinates (\ref{KS-coors}) defined in the domains $-\infty<U<\infty$ and $-\infty<V<\infty$ cover the entire maximally extended Schwarzschild spacetime.

In order to draw the Penrose diagram shown in Fig.~\ref{SingleBH}, one needs to introduce new null coordinates ${\bar U}$ and ${\bar V}$ such that
\begin{eqnarray}
{\bar U}:=\arctan U, \qquad {\bar V}:=\arctan V, \label{barUV-def}
\end{eqnarray}
of which domains are $-\pi/2<{\bar U}<\pi/2$ and $-\pi/2<{\bar V}<\pi/2$.
Now the Schwarzschild spacetime is embedded in finite domains of ${\bar U}$ and ${\bar V}$ and the line element on $(M^2,g_{AB})$ is written as $\D s_{2}^2=(\cos {\bar U} \cos {\bar V})^{-2} \D{\tilde s}_{{2}}^2$, where
\begin{align}
\D{\tilde s}_{{2}}^2:= - \frac{32M^3}{r} e^{-r/(2M)} \D {\bar U} \D {\bar V}.\label{Sc5}
\end{align}
The spacetime with the metric (\ref{Sc5}) provides the Penrose diagram in Fig.~\ref{SingleBH}, which is a conformal completion of the maximally extended Schwarzschild spacetime by attaching the boundaries at ${\bar U}=\pm\pi/2$ and ${\bar V}=\pm\pi/2$.

\noindent
{\bf Isotropic coordinates}

With a new radial coordinate $\sigma$ defined by
\begin{eqnarray}
r=\left(1+\frac{M}{2\sigma}\right)^2\sigma, \label{r-rho}
\end{eqnarray}
the Schwarzschild metric~(\ref{Sc1}) is written in the isotropic coordinates as
\begin{eqnarray}
\D s^2&=&-\biggl(\frac{2\sigma-M}{2\sigma+M}\biggl)^2\D t^2+\left(1+\frac{M}{2\sigma}\right)^4\left(\D \sigma^2+\sigma^2\D \Omega^2\right), \label{Sch-iso}
\end{eqnarray}
where $\sigma=M/2$ is a coordinate singularity.
Since the areal radius $r=r(\sigma)$ given by Eq.~(\ref{r-rho}) takes a minimum value $r=2M$ at the ``throat'' $\sigma=M/2$, the coordinates (\ref{Sch-iso}) cover only regions with $r>2M$, which are the regions I and III in Fig.~\ref{SingleBH}, and $\sigma=0$ and $\sigma=\infty$ correspond to two distinct spacelike infinities.

\noindent
{\bf Painlev\'{e}-Gullstrand coordinates}

\begin{figure}[htbp]
\begin{center}
\includegraphics[width=0.7\linewidth]{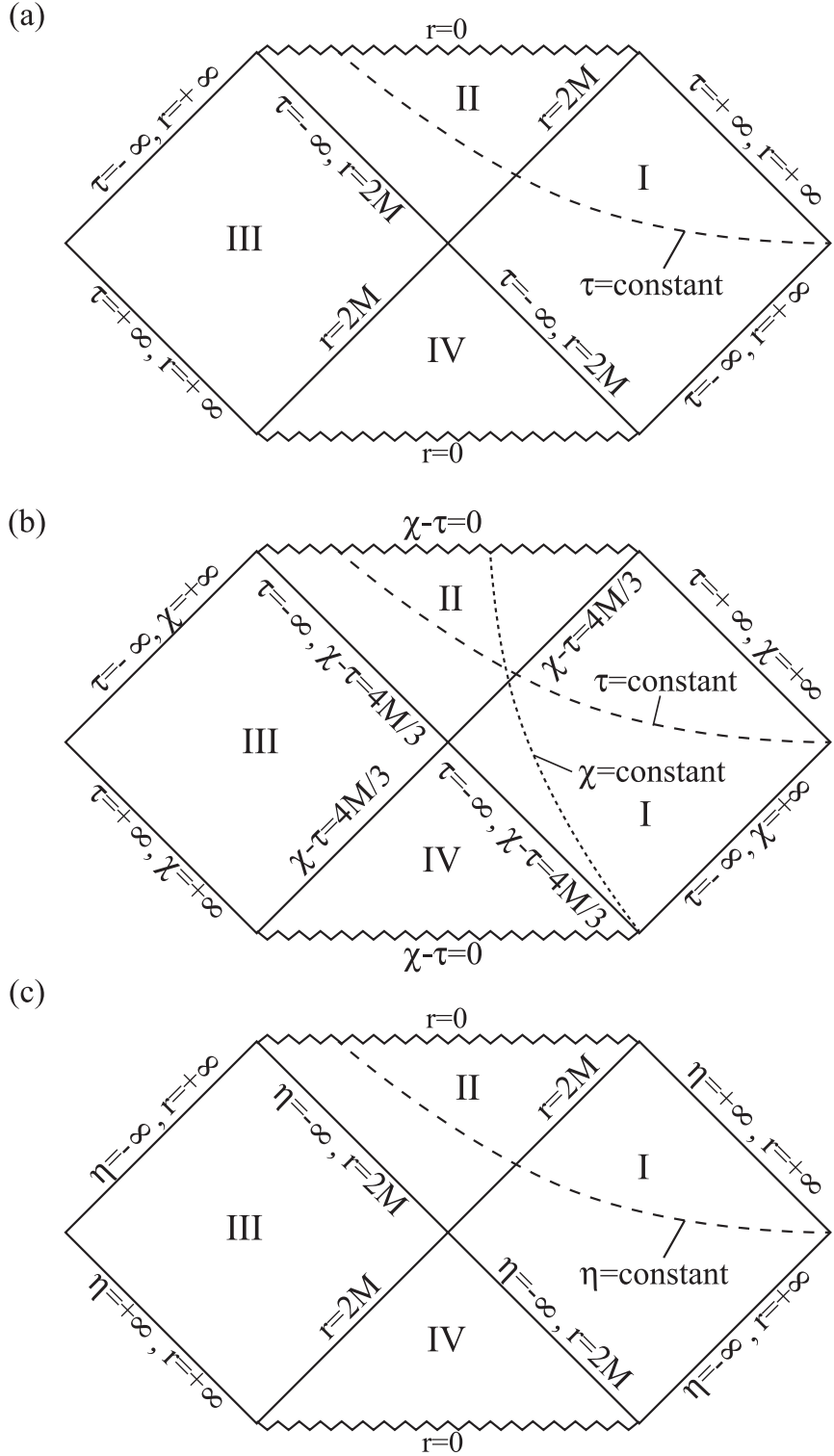}
\caption{\label{Schwarzschild-3coordinates} 
(a) The Painlev\'{e}-Gullstrand coordinates (\ref{Sch-P}), (b) the Lema\^{\i}tre coordinates (\ref{Sch-L}), and (c) the Kerr-Schild coordinates (\ref{Sch-KS})
in the Penrose diagrams of the maximally extended Schwarzschild spacetime.}
\end{center}
\end{figure}

With a new coordinate $\tau$ defined by
\begin{equation}
\tau := t+2\sqrt{2Mr}\biggl\{1-{\frac12\sqrt{\frac{2M}{r}}\ln \left|\frac{\sqrt{2M/r}+1}{\sqrt{2M/r}-1}\right|}\biggl\},\label{def:tau-PG}
\end{equation}
which satisfies $\D\tau=\D t+\sqrt{2M/r}\D r/(1-2M/r)$, the Schwarzschild metric~(\ref{Sc1}) is written in the Painlev\'{e}-Gullstrand coordinates as
\begin{align}
\D s^2=&-\biggl(1-\frac{2M}{r}\biggl)\D \tau^2+2\sqrt{\frac{2M}{r}}\D \tau \D r+\D r^2+r^2\D \Omega^2 \nonumber \\
=&-\D \tau^2+\biggl(\sqrt{\frac{2M}{r}}\D \tau +\D r\biggl)^2+r^2\D \Omega^2. \label{Sch-P}
\end{align}
Non-zero components of the inverse metric on $(M^2,g_{AB})$ are
\begin{align}
g^{\tau\tau}=-1,\qquad g^{rr}=1-\frac{2M}{r}, \qquad g^{\tau r}(=g^{r\tau})=\sqrt{\frac{2M}{r}}.
\end{align}
It is noted that $\tau=$constant is a spacelike hypersurface but $\partial/\partial \tau$ is timelike (spacelike) in a region with $r>(<)2M$.
Since the metric and its inverse are regular at $r=2M$, the domains of $\tau$ and $r$ are $-\infty<\tau<\infty$ and $0<r<\infty$, respectively.
As a result, the Painlev\'{e}-Gullstrand coordinates (\ref{Sch-P}) cover the regions I$+$II or III$+$IV in Figs.~\ref{SingleBH} and~\ref{Schwarzschild-3coordinates}(a).

\noindent
{\bf Lema\^{\i}tre coordinates}

By a coordinate transformation
\begin{equation}
r=(2M)^{1/3}\left[\frac32(\chi-\tau)\right]^{2/3} \label{r-Lemaitre}
\end{equation}
from the Painlev\'{e}-Gullstrand coordinates (\ref{Sch-P}), one obtains the Schwarzschild metric in the Lema\^{\i}tre coordinates;
\begin{eqnarray}
\D s^2=-\D \tau^2+(2M)^{2/3}\biggl(\frac{\D \chi^2}{\left[\frac32(\chi-\tau)\right]^{2/3}}+\left[\frac32(\chi-\tau)\right]^{4/3}\D \Omega^2\biggl). \label{Sch-L}
\end{eqnarray}
$\tau$ is a timelike coordinate and $\chi$ is a spacelike coordinate everywhere and their domains are $-\infty<\tau<\infty$ and $\chi>\tau$.
The hypersurface-orthogonal Killing vector $\xi^\mu$ generating staticity in an untrapped region is given by
\begin{align}
\xi^\mu\frac{\partial}{\partial x^\mu}=\frac{\partial}{\partial \tau}+\frac{\partial}{\partial \chi},
\end{align}
of which squared norm is
\begin{align}
\xi_\mu \xi^\mu=-1+\left[\frac{3(\chi-\tau)}{4M}\right]^{-2/3}.
\end{align}
The areal radius $r=r(\tau,\chi)$ given by Eq.~(\ref{r-Lemaitre}) is a monotonically increasing function of $\chi$ and a monotonically decreasing function of $\tau$.
A curvature singularity $r=0$ and the event horizon $r=2M$ correspond to $\tau=\chi$ and $\chi-\tau=4M/3$, respectively.
Since the metric and its inverse are regular at $r=2M$, the Lema\^{\i}tre coordinates (\ref{Sch-L}) cover the region I$+$II or III$+$IV in Figs.~\ref{SingleBH} and~\ref{Schwarzschild-3coordinates}(b).

\noindent
{\bf Kerr-Schild coordinates}

With a new coordinate $\eta$ defined by
\begin{align}
\eta:=v-r, \label{def-eta-KS}
\end{align}
the Schwarzschild metric in the ingoing-null Eddington-Finkelstein coordinates (\ref{Sc3}) is written in the Kerr-Schild coordinates\footnote{Historically, the coordinates~(\ref{Sch-KS}) in the Schwarzschild spacetime were introduced by Eddington already in 1924~\cite{Eddington1924} much earlier than Kerr and Schild introduced their coordinates in the Kerr spacetime in 1956~\cite{KerrSchild1956}. Nevertheless, we refer to the coordinates (\ref{Sch-KS}) as the Kerr-Schild coordinates to distinguish them from the Eddington-Finkelstein coordinates.} as
\begin{equation}
\D s^2=-\D \eta^2+\D r^2+r^2\D \Omega^2+\frac{2M}{r}(\D \eta+\D r)^2. \label{Sch-KS}
\end{equation}
Non-zero components of the inverse metric on $(M^2,g_{AB})$ are
\begin{align}
g^{\eta\eta}=-\biggl(1+\frac{2M}{r}\biggl),\qquad g^{rr}=1-\frac{2M}{r}, \qquad g^{\eta r}(=g^{r\eta})=\frac{2M}{r},
\end{align}
and hence $\eta=$constant is a spacelike hypersurface.
Since the metric and its inverse are regular at $r=2M$, the domains of $\eta$ and $r$ are $-\infty<\eta<\infty$ and $0<r<\infty$.
As a result, the Kerr-Schild coordinates (\ref{Sch-KS}) cover the region I$+$II or III$+$IV in Figs.~\ref{SingleBH} and~\ref{Schwarzschild-3coordinates}(c).

\subsection{Conformally Schwarzschild spacetime}

In subsequent sections, we will study a variety of spacetimes $({M}^4,{g}_{\mu\nu})$ which are conformally related to the Schwarzschild spacetime $({\bar M}^4,{\bar g}_{\mu\nu})$ as ${g}_{\mu\nu}=\Omega^{2}{\bar g}_{\mu\nu}$.
Although a conformal transformation does not change light-cone structure, it does change the nature of the coordinate boundaries in the Penrose diagram.
For example, in a spacetime which is conformally related to the Schwarzschild spacetime in the standard Schwarzschild coordinates (\ref{Sc1}), $r=2M$ in Fig.~\ref{SingleBH} can be a curvature singularity or null infinity depending on the form of $\Omega$.
Similarly, a future null infinity in Fig.~\ref{SingleBH} can be an extendable boundary in the coordinate system on $({M}^4,{g}_{\mu\nu})$.
Since the conformal factor $\Omega^2$ may introduce a new curvature singularity or infinity in $({M}^4,{g}_{\mu\nu})$, its global structure may be quite different from the Schwarzschild spacetime $({\bar M}^4,{\bar g}_{\mu\nu})$.

In addition, the conformal factor $\Omega^2$ generally introduces a different matter field in $({M}^4,{g}_{\mu\nu})$.
With the vanishing Einstein tensor ${\bar G}_{\mu\nu}\equiv 0$ of $({\bar M}^4,{\bar g}_{\mu\nu})$, the Einstein tensor ${G}_{\mu\nu}$ of $({M}^4,{g}_{\mu\nu})$ is given by
\begin{align}
{G}_{\mu\nu}=-2\Omega^{-1}\nabla_\mu \nabla_\nu\Omega+{g}_{\mu\nu}\left[2\Omega^{-1}\nabla^2\Omega-3\Omega^{-2}(\nabla \Omega)^2\right],\label{conformal-G}
\end{align}
where $\nabla_\mu$ is a covariant derivative on $({M}^4,{g}_{\mu\nu})$~\cite{wald1983}.
In general relativity, the right-hand side of Eq.~(\ref{conformal-G}) is identical to the energy-momentum tensor $8\pi T_{\mu\nu}$ in the spacetime $({M}^4,{g}_{\mu\nu})$.

Furthermore, the conformal factor may change the spacetime symmetries.
If $({\bar M}^4,{\bar g}_{\mu\nu})$ admits a Killing vector ${\xi}^\mu$ satisfying a Killing equation ${\cal L}_\xi {\bar g}_{\mu\nu}=0$, one obtains
\begin{align}
{\cal L}_\xi g_{\mu\nu}=2\Omega^{-1}g_{\mu\nu}\xi^\rho\nabla_\rho\Omega.\label{CKeq}
\end{align}
Hence, the conformally related spacetime $({M}^4,{g}_{\mu\nu})$ admits a conformal Killing vector $\xi^\mu$ satisfying a conformal Killing equation ${\cal L}_\xi g_{\mu\nu}=2\psi g_{\mu\nu}$, where $\psi$ is given by 
\begin{align}
\psi=\Omega^{-1}\xi^\rho\nabla_\rho\Omega.
\end{align}
Then, a conformal Killing horizon $\Sigma$ is defined in $({M}^4,{g}_{\mu\nu})$ in a parallel way to a Killing horizon as a null hypersurface where the conformal Killing vector $\xi^\mu$ becomes null~\cite{DyerHonig1979,SultanaDyer2004}.

With a suitable conformal factor $\Omega^2$, a conformal Killing horizon $\Sigma$ in the spacetime $({M}^4,{g}_{\mu\nu})$ and a Killing horizon in the Schwarzschild black-hole spacetime $({\bar M}^4,{\bar g}_{\mu\nu})$ may coincide with the same event horizon.
If a black-hole spacetime $({M}^4,{g}_{\mu\nu})$ is static and asymptotically flat, the Hawking radiations of a conformally coupled scalar field from these two conformally related black holes are the same.
For this reason, Jacobson and Kang argued that the surface gravity and temperature of a black hole, which characterizes the Hawking radiation, should be conformally invariant~\cite{Jacobson:1993pf}.
Then, they defined the surface gravity $\kappa$ and temperature $T_{\rm JKSD}$ on $\Sigma$ in terms of the conformal Killing vector $\xi^\mu$ by
\begin{align}
\nabla_\mu(\xi^\nu\xi_\nu)|_\Sigma=-2\kappa\xi_\mu|_\Sigma,\qquad T_{\rm JKSD}:=\frac{\kappa}{2\pi}.\label{T-CK}
\end{align}
On the other hand, Sultana and Dyer defined the surface gravity $\kappa_{\rm SD}$ by
\begin{align}
\xi^\nu\nabla_\nu \xi_\mu|_\Sigma=\kappa_{\rm SD}\xi_\mu|_\Sigma \label{kappa-SD}
\end{align}
in a different approach~\cite{SultanaDyer2004}.
Although both $\kappa$ and $\kappa_{\rm SD}$ reduce to the same surface gravity if $\xi^\mu$ is a Killing vector, $\kappa_{\rm SD}$ is not conformally invariant and satisfies $\kappa=\kappa_{\rm SD}-2\psi$~\cite{Jacobson:1993pf}\footnote{Another different definition of surface gravity has been proposed in~\cite{DyerHonig1979}.}.
Nevertheless, Sultana and Dyer~\cite{SultanaDyer2004} defined the temperature by $T_{\rm SD}:=(\kappa_{\rm SD}-2\psi|_\Sigma)/2\pi$, which is identical to $T_{\rm JKSD}$ and conformally invariant.
Since $T_{\rm JKSD}$ is conformally invariant, with the same normalization of $\xi^\mu$ for the Schwarzschild black hole ($\Omega\equiv 1$), one obtains $T_{\rm JKSD}=1/(4M)$ for a conformally Schwarzschild black hole $({M}^4,{g}_{\mu\nu})$.
However, it was shown that the effective temperature of the Sultana-Dyer cosmological black hole evaluated from the Hawking radiation is time-dependent and modified from $T_{\rm JKSD}$~\cite{Saida:2007ru}.

As an alternative definition of a dynamical black hole, one could consider a future-outer trapping horizon.
In terms of the Kodama vector $K^\mu(\partial/\partial x^\mu)=K^A(\partial/\partial y^A)$ with Eq.~(\ref{kodamavector}), the surface gravity $\kappa_{\rm TH}$ and temperature $T_{\rm TH}$ on an outer or degenerate trapping horizon $\Sigma$ are defined by~\cite{Hayward:1997jp}
\begin{align}
K^\nu\nabla_{[\nu}K_{\mu]}|_\Sigma=\kappa_{\rm TH}K_\mu|_\Sigma,\qquad T_{\rm TH}:=\frac{\kappa_{\rm TH}}{2\pi}.\label{T-TH}
\end{align}
In the static case, $K^\mu$ and $\kappa_{\rm TH}$ reduce to a hypersurface-orthogonal Killing vector $\xi^\mu$ and the surface gravity on a Killing horizon, respectively.
It has been reported that the Hawking temperature of any future outer trapping horizon in a spherically symmetric spacetime derived by a Hamilton-Jacobi variant of the Parikh-Wilczek tunneling method coincides with $T_{\rm TH}$~\cite{Hayward:2008jq}.

\subsection{Unsuccessful models of a cosmological black hole}
\label{sec:fail}

In the present paper, we will investigate various conformally Schwarzschild spacetimes which are asymptotically flat FLRW universe filled by a perfect fluid obeying a linear equation state $p=w\rho$ with $w>-1/3$ to seek for cosmological black-hole solutions. 
In particular, we will consider the Thakurta spacetime (\ref{Thakurta1}), the McClure-Dyer spacetime (\ref{MD}), the Sultana-Dyer class of spacetimes (\ref{metricSD}), and the Culetu spacetime (\ref{PG-c}).
Before moving to the analyses of the latter two spacetimes in the subsequent sections, here we show that the Thakurta spacetime and the McClure-Dyer spacetime are identical and they do not describe a cosmological black hole based on the previous works~\cite{Mello:2016irl,Harada:2021xze}.

Actually, in spite that the Thakurta spacetime (or equivalently the McClure-Dyer spacetime) is distinct from other two spacetimes unless the conformal factor $a^2$ is non-constant, it has been misidentified with the Sultana-Dyer class of spacetimes by incorrect coordinate transformations disregarding the integrability conditions in some papers~\cite{Faraoni:2007es,Faraoni:2007gq,Faraoni:2009uy}.
When a coordinate transformation $y =y({\bar y})$ on $(M^2,g_{AB})$ is defined in terms of differential displacements such that $\D y=F_A({\bar y})\D \bar{y}^A$, the functions $F_A({\bar y})$ must satisfy an integrability condition $\partial F_0/\partial {\bar y}^1=\partial F_1/\partial {\bar y}^0$.

\subsubsection{Thakurta class}
\label{sec:Thakurta}

\begin{figure}[htbp]
\begin{center}
\includegraphics[width=0.7\linewidth]{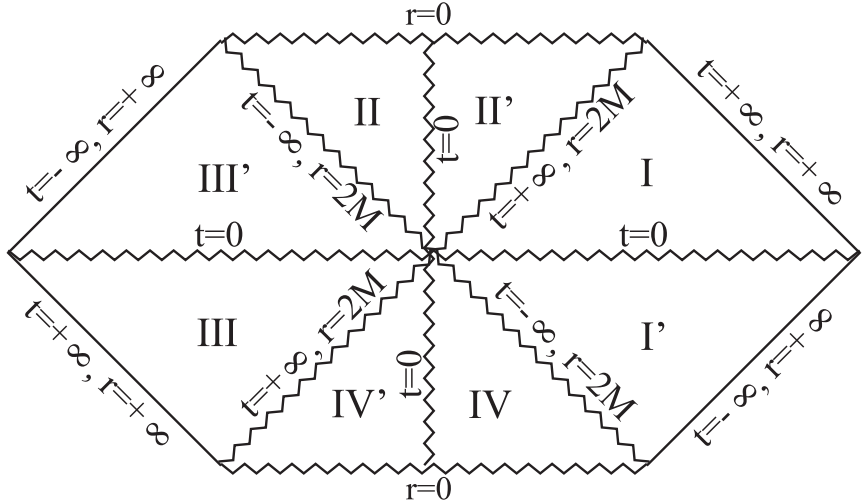}
\caption{\label{Thakurta} A Penrose diagram of the maximally extended Thakurta spacetime (\ref{Thakurta1}) with $a(t)=a_0|t|^\alpha~(\alpha>0)$. Regions with different labels are causally disconnected and describe distinct spacetimes.}
\end{center}
\end{figure}

The Thakurta spacetime~\cite{Thakurta1983} is a conformally Schwarzschild spacetime constructed with the Schwarzschild coordinates (\ref{Sc1}) given by
\begin{align}
\label{Thakurta1}
\begin{aligned}
&\D s^2=a(t)^2\left[-f(r)\D t^2+f(r)^{-1}\D r^2+r^2\D\Omega^2\right],\\
&f(r):=1-\frac{2M}{r},
\end{aligned}
\end{align}
which is asymptotic to the flat FLRW spacetime as $r\to \infty$.
The global structure of the Thakurta spacetime with $M>0$ and $a(t)=a_0t^\alpha~(\alpha>0)$ has been clarified in~\cite{Mello:2016irl,Harada:2021xze}.
In this case, $t=0$ and $r=2M$ are curvature singularities and the maximally extended spacetime given in the domains of $0<t<\infty$ and $2M<r<\infty$ does not admit neither event horizon nor future outer trapping horizon.
As the Penrose diagram drawn in Fig.~\ref{Thakurta} shows, the Thakurta spacetime does not describe a cosmological black hole.

\subsubsection{McClure-Dyer class}

The McClure-Dyer spacetime~\cite{McClure:2006kg} is a conformally Schwarzschild spacetime constructed with the isotropic coordinates (\ref{Sch-iso}) given by
\begin{eqnarray}
\D s^2&=&a(t)^2\biggl[-\biggl(\frac{2\sigma-M}{2\sigma+M}\biggl)^2\D t^2+\left(1+\frac{M}{2\sigma}\right)^4(\D \sigma^2+\sigma^2\D \Omega^2)\biggl]. \label{MD}
\end{eqnarray}
In fact, the McClure-Dyer spacetime is obtained from the Thakurta spacetime (\ref{Thakurta1}) by a coordinate transformation (\ref{r-rho}).
Therefore, the coordinates (\ref{MD}) cover the regions of $r>2M$ in the Thakurta spacetime corresponding to I, I', III, and III' in Fig.~\ref{Thakurta}.
As a result, the McClure-Dyer spacetime does not describe a cosmological black hole either.

\section{Sultana-Dyer class}
\label{sec:SD}

In this section, we investigate the following conformally Schwarzschild spacetime constructed with the Kerr-Schild coordinates:
\begin{equation}
\D s^2=a(\eta)^2\biggl[-\D \eta^2+\D r^2+r^2\D \Omega^2+\frac{2M}{r}(\D \eta+\D r)^2\biggl], \label{metricSD}
\end{equation}
which is asymptotic to the flat FLRW spacetime as $r\to \infty$.
Non-zero components of the inverse metric are given by 
\begin{align}
\begin{aligned}
&g^{\eta\eta}=-\frac{1}{a^2}\biggl(1+\frac{2M}{r}\biggl),\qquad g^{rr}=\frac{1}{a^2}\biggl(1-\frac{2M}{r}\biggl),\\
&g^{\eta r}(=g^{r\eta})=\frac{2M}{a^2r},\qquad g^{\theta\theta}=g^{\phi\phi}\sin^2\theta=\frac{1}{a^2r^2}.
\end{aligned}
\end{align}
Sultana and Dyer studied the spacetime (\ref{metricSD}) in detail with $a(\eta)=\eta^2$~\cite{Sultana:2005tp}.
In this section, we assume $M>0$ and study a more general case $a(\eta) = a_{0}|\eta|^{\alpha}$, where $a_{0}$ and $\alpha$ are positive constants.
In the spacetime, $\eta$ is a timelike coordinate everywhere and we define the future direction by increasing $\eta$.

\subsection{Global structure}

Under our assumptions, the spacetime with the metric (\ref{metricSD}) is analytic except at $\eta=0$ and $r=0$.
In fact, both $\eta=0$ and $r=0$ are curvature singularities as the following Ricci scalar ${\cal R}$ blows up:
\begin{align}
{\cal R} = \frac{6}{r^2 a^2}\biggl\{r^2\frac{\ddot{a}}{a} + 2M\biggl(r\frac{\ddot{a}}{a}- \frac{\dot{a}}{a}\biggl)\biggl\}= \frac{6\alpha\{(\alpha -1)r(r+2M) - 2M\eta\}}{a_{0}^2 r^2 |\eta|^{2(1+\alpha)}}.\label{Ricci-s-SD}
\end{align}
In the domain $\eta\in(0,\infty)$ ($\eta\in(-\infty,0)$), the spacetime approaches as $r\to \infty$ the asymptotically flat Friedmann expanding (collapsing) universe with a perfect fluid that obeys an equation of state $p=w\rho$, where $w$ satisfies $\alpha = 2/(1 + 3w)$, so that $\alpha>0$ is equivalent to $w > -1/3$.

Since the spacetime admits a hypersurface-orthogonal conformal Killing vector $\xi^\mu=(1,0,0,0)$ satisfying ${\cal L}_\xi g_{\mu\nu}=2({\dot a}/a)g_{\mu\nu}$, there is a conserved quantity $C:=-\xi_\mu {\bar k}^\mu$ along a null geodesic $\gamma$ with its tangent vector ${\bar k}^\mu$.
If $\gamma$ is a future-directed radial null geodesic, it is described by $x^\mu=(\eta(\lambda), r(\lambda),0,0)$, where $\lambda$ is an affine parameter along $\gamma$.
Then, we have ${\bar k}^\mu=(\D \eta/\D\lambda, \D r/\D\lambda,0,0)$ and 
\begin{equation}
C=a(\eta)^2\biggl[\biggl(1-\frac{2M}{r}\biggl){\bar k}^0-\frac{2M}{r}{\bar k}^1\biggl].\label{C-SD}
\end{equation}
By the null condition $\D s^2=0$ for $\gamma$, we obtain
\begin{align}
\biggl(1+\frac{2M}{r}\biggl){\bar k}^1=\biggl(1-\frac{2M}{r}\biggl){\bar k}^0 \label{k-out-SD}
\end{align}
for outgoing $\gamma$ and 
\begin{align}
{\bar k}^1=-{\bar k}^0 \label{k-in-SD}
\end{align}
for ingoing $\gamma$.
Equation~(\ref{k-out-SD}) is integrated to give
\begin{align}
\eta-\eta_0=&r+4M\ln\biggl|\frac{r}{2M}-1\biggl|,\label{k-out-int-SD}
\end{align}
where $\eta_0$ is an integration constant.
Equation~(\ref{k-out-int-SD}) shows $\eta\to \infty$ as $r\to \infty$ and $\eta\to -\infty$ as $r\to 2M$, as seen in Fig.~\ref{Schwarzschild-3coordinates}(c).
Equations~(\ref{C-SD}) and (\ref{k-out-SD}) give
\begin{equation}
\frac{C}{a(\eta)^2}={\bar k}^1.\label{C-int-SD}
\end{equation}
With $a(\eta)=a_0|\eta|^\alpha$ and Eq.~(\ref{k-out-int-SD}), Eq.~(\ref{C-int-SD}) is integrated to give
\begin{equation}
\frac{C}{a_0^2}(\lambda-\lambda_0)=\int^r\left|\eta_0+{\bar r}+4M\ln\biggl|\frac{{\bar r}}{2M}-1\biggl|\right|^{2\alpha}\D {\bar r},\label{F-int-SD}
\end{equation}
where $\lambda_0$ is an integration constant.
Since the affine parameter $\lambda$ diverges as $r\to\infty$ along a future-directed outgoing radial null geodesic $\gamma$, $(\eta,r)\to (\infty,\infty)$ is a future null infinity.
In contrast, the right-hand side of Eq.~(\ref{F-int-SD}) is shown to be finite as $r\to 2M$.
For any $\epsilon$ with $0<\epsilon<1/(2\alpha)$, there exists $\delta>0$ such that 
\begin{equation}
\left|\eta_0+{\bar r}+4M\ln\biggl|\frac{{\bar r}}{2M}-1\biggl|\right|
<4M\left(\frac{{\bar r}}{2M}-1\right)^{-\epsilon}
\end{equation}
for $2M-\delta <\bar{r}<2M$. Then, it immediately follows that the integral of the right-hand side of Eq.~(\ref{F-int-SD}) is bounded from above.
Since the affine parameter $\lambda$ is finite, $(\eta,r)\to (-\infty,2M)$ is an extendable boundary if it is regular.
Although the Ricci scalar and the Kretschmann scalar ${\cal R}_{\mu\nu\rho\sigma}{\cal R}^{\mu\nu\rho\sigma}$ are finite as $(\eta,r)\to (-\infty,2M)$ along $\gamma$, it could also be a parallelly propagated (p.p.) curvature singularity, which is defined by the fact that some component of the Riemann tensor in the parallelly transported frame along a geodesic blows up~\cite{he1973}.

On the other hand, Eq~(\ref{k-in-SD}) is integrated to give $r=-(\eta-\eta_0)$, which shows $\eta\to -\infty$ as $r\to \infty$, as seen in Fig.~\ref{Schwarzschild-3coordinates}(c).
Equations~(\ref{C-SD}) and (\ref{k-in-SD}) give
\begin{equation}
\frac{C}{a(\eta)^2}={\bar k}^0.\label{C-int-in-SD}
\end{equation}
With $a(\eta)=a_0|\eta|^\alpha$, Eq.~(\ref{C-int-in-SD}) is integrated to give
\begin{equation}
\frac{C}{a_0^2}(\lambda-\lambda_0)=\int^\eta |{\bar \eta}|^{2\alpha}\D {\bar \eta}.
\end{equation}
Since $|\lambda|\to \infty$ holds as $\eta\to -\infty$ (and hence $r\to \infty$) along a future-directed ingoing radial null geodesic, $(\eta,r)\to (-\infty,\infty)$ is a past null infinity.

\begin{figure}[htbp]
\begin{center}
\includegraphics[width=0.7\linewidth]{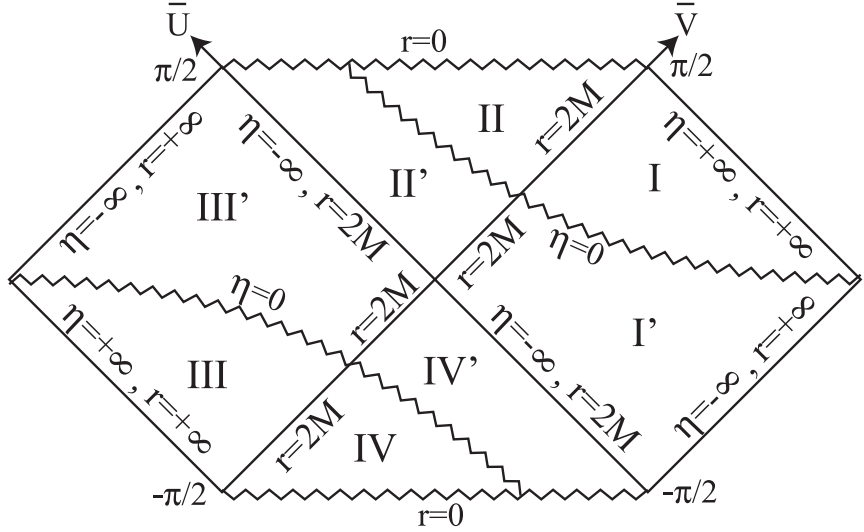}
\caption{\label{Sultana-Dyer} A Penrose diagram of the maximally extended Sultana-Dyer spacetime (\ref{metricSD}) with $a(\eta)=a_0|\eta|^\alpha~(\alpha>0)$. The portion consisting of I and II describes a cosmological black hole. If the coordinate boundary $(\eta,r)\to (-\infty,2M)$ is regular, it is extendable. }
\end{center}
\end{figure}

As a result, the Penrose diagram of the spacetime 
(\ref{metricSD}) with $a(\eta)=a_0|\eta|^\alpha~(\alpha>0)$ is drawn as in Fig.~\ref{Sultana-Dyer}.
It is clear that a maximally symmetric spacetime given in the domains $0<\eta<\infty$ and $0<r<\infty$, which corresponds to the portion consisting of I and II, describes an asymptotically flat FLRW cosmological black hole with the event horizon at $r=2M$.
While $r=0$ corresponds to a black-hole singularity, $\eta=0$ corresponds to a big-bang singularity.
Its time-reversal spacetime consisting of III and IV, with the future direction defined by decreasing $\eta$, describes a white hole in the flat FLRW collapsing universe.
If the coordinate boundary $(\eta,r)\to (-\infty,2M)$ is regular and then extendable, the maximally extended spacetime consisting of I', II', III', and IV' also describes a cosmological black hole.
Even in that case, since this portion is not covered by a single coordinate system (\ref{metricSD}), the regular metric at the event horizon $(\eta,r)\to (-\infty,2M)$ could be non-analytic and allow for lower differentiability.
Hereafter, we will focus on the cosmological black hole corresponding to I$+$II.

\subsection{Matter fields}

Now let us identify the matter field to give the corresponding energy-momentum tensor $T_{\mu\nu}(=G_{\mu}/8\pi)$ in the cosmological black-hole spacetime in the domains $0<\eta<\infty$ and $0<r<\infty$.
Non-zero components of the Einstein tensor for the metric (\ref{metricSD}) are given by
\begin{align}
G^{00} &=\frac{\dot{a}}{r^3 a^5} \biggl\{-4M(r + 3M) + 3r(r + 2M)^2 \frac{\dot{a}}{a}\biggl\}, \label{Einstein-SD0}\\
G^{11} &= \frac{1}{r^3 a^4}\biggl\{r(r^2 + 12M^2)\frac{\dot{a}^2}{a^2} + 4M(2r - 3M)\frac{\dot{a}}{a} - 2r^3 \frac{\ddot{a}}{a}\biggl\},\label{Einstein-SD1}\\
G^{01} &= -\frac{2M\dot{a}}{r^3 a^5}\biggl\{(r - 6M)+ 3r(r + 2M)\frac{\dot{a}}{a} \biggl\},\label{Einstein-SD2}\\
G^{22} &= \frac{G^{33}}{\sin^2\theta}=\frac{r + 2M}{r^3 a^4}\biggl(\frac{\dot{a}^2}{a^2} - 2\frac{\ddot{a}}{a}\biggl). \label{Einstein-SD3}
\end{align}
Adopting the orthonormal basis one-forms
\begin{align}
E_\mu^{(0)}\D x^\mu=&-a\biggl[\biggl(1-\frac{M}{r}\biggl)\D \eta-\frac{M}{r}\D r\biggl],\label{E0-SD}\\
E_\mu^{(1)}\D x^\mu=&a\biggl[\frac{M}{r}\D \eta + \biggl(1+\frac{M}{r}\biggl)\D r\biggl],\\
E_\mu^{(2)}\D x^\mu=&ar\D \theta, \qquad E_\mu^{(3)}\D x^\mu=ar\sin\theta \D \phi,\label{E2-SD}
\end{align}
we obtain
\begin{align}
G^{(0)(0)}=&8\pi\biggl(\rho_{F}+\mu+\frac12\Omega\biggl),\qquad G^{(1)(1)}=8\pi\biggl(p_{F}-\mu+\frac12\Omega\biggl),\label{G00-SD}\\
G^{(0)(1)}=&G^{(1)(0)}=4\pi\Omega,\qquad G^{(2)(2)}=G^{(3)(3)}=8\pi(p_{F}+P),\label{G22-SD}
\end{align}
where $\rho_{F}$, $p_{F}$, $\mu$, $P$, and $\Omega$ are given by 
\begin{align}
&8\pi \rho_{F}=\frac{3{\dot a}^2}{a^4},\qquad 8\pi p_{F}=\frac{{\dot a}^2}{a^4}-\frac{2{\ddot a}}{a^3},\label{rho-p-F-def}\\
&8\pi\mu=\frac{2M}{r^2}\biggl(r\frac{{\dot a}^2}{a^4}+r\frac{{\ddot a}}{a^3}-\frac{3{\dot a}}{a^3}\biggl),\qquad 8\pi P=\frac{2M}{r}\biggl(\frac{{\dot a}^2}{a^4}-\frac{2{\ddot a}}{a^3}\biggl),\label{mu-P-def}\\
&4\pi \Omega=\frac{2M}{r^2}\biggl\{4\pi(r+M)(\rho_F+p_F)+\frac{{\dot a}}{a^3}\biggl\}.\label{emt-SD2-last}
\end{align}
If the NEC is satisfied in the asymptotically flat FLRW region, $\rho_F+p_F\ge 0$ holds.
Under this condition together with $M\ge 0$ and ${\dot a}\ge 0$, we can show the following proposition.
\begin{Prop}
\label{Prop:EC-SD}
Suppose $\rho_F+p_F\ge 0$ in the Sultana-Dyer class of spacetimes (\ref{metricSD}).
Then, the corresponding energy-momentum tensor in general relativity is of the Hawking-Ellis type I if $M\ge 0$ and ${\dot a}\ge 0$ hold.
In addition, if $M> 0$ and ${\dot a}> 0$ hold, it violates all the standard energy conditions near a curvature singularity at $r=0$.
\end{Prop}
{\it Proof:}
Equations~(\ref{G00-SD})--(\ref{emt-SD2-last}) give
\begin{align}
&(G^{(0)(0)}+G^{(1)(1)})^2- 4(G^{(0)(1)})^2=(8\pi)^2(\rho_F+p_F)(\rho_F+p_F+2\Omega),\label{SD-general1}\\
&G^{(0)(0)}+G^{(1)(1)}=8\pi(\rho_F+p_F)\biggl\{1+\frac{2M}{r^2}(r+M)\biggl\}+\frac{4M}{r^2}\frac{{\dot a}}{a^3},\label{SD-general1}
\end{align}
which are non-negative under $\rho_F+p_F\ge 0$, $M\ge 0$, and ${\dot a}\ge 0$.
Hence, the corresponding energy-momentum tensor is of the Hawking-Ellis type I by Lemma~\ref{Prop:HE-type} and the NEC is equivalent to ${\cal T}_{\rm N}\ge 0$ by Proposition~\ref{Prop:EC-criteriaI}, where
\begin{align}
&{\cal T}_{\rm N}:=(G^{(0)(0)}-G^{(1)(1)}+2G^{(2)(2)})+\sqrt{(G^{(0)(0)}+G^{(1)(1)})^2-4(G^{(0)(1)})^2} \nonumber \\
&~~~~=8\pi\biggl\{\rho_F+p_F+2(\mu+P)+\sqrt{(\rho_F+p_F)(\rho_F+p_F+2\Omega)}\biggl\}.\label{def-TN}
\end{align}
The first expression of the Ricci scalar (\ref{Ricci-s-SD}) shows that $r=0$ is a curvature singularity if ${\dot a}\ne 0$ and $M\ne 0$ are satisfied.
Under the assumptions $\rho_F+p_F\ge 0$, $M> 0$, and ${\dot a}> 0$, we obtain 
\begin{align}
&\lim_{r\to 0}{\cal T}_{\rm N}\simeq -\frac{4M}{r^2}\times \frac{3{\dot a}}{a^3}\to -\infty,
\end{align}
so that all the standard energy conditions are violated near $r=0$.
\qed

Proposition~\ref{Prop:EC-SD} shows that, under the standard energy conditions, the Sultana-Dyer class of spacetimes cannot be a model of a cosmological black hole near the curvature singularity $r=0$ in the framework of general relativity.
Now let us identify the regions where the energy conditions are violated in the domain of $\eta\ge 0$ for $a(\eta)=a_0 \eta^\alpha$ ($\alpha>0$).
Since Eq.~(\ref{def-TN}) gives
\begin{align}
{\cal T}_{\rm N}=&\frac{2\alpha(\alpha+1)}{\eta^2a^2}\biggl[1+\frac{2M}{r^2}\biggl(r-\frac{3\eta}{\alpha+1}\biggl)+\sqrt{\frac{(\alpha+1)(r+2M)^2+4M\eta}{(\alpha+1)r^2}}\biggl],
\end{align}
an inequality ${\cal T}_{\rm N}\ge 0$ equivalent to the NEC is satisfied if and only if $0<\eta\le \eta_{\rm N}(r)$ holds, where
\begin{align}
\eta_{\rm N}(r):=\frac{2(\alpha+1)r(2r+3M)}{9M}.\label{NEC-SD}
\end{align}

To check the WEC, we compute
\begin{align}
{\cal T}_{\rm W}:=&(G^{(0)(0)}-G^{(1)(1)})+\sqrt{(G^{(0)(0)}+G^{(1)(1)})^2-4(G^{(0)(1)})^2} \nonumber \\
=&\frac{2\alpha(\alpha+1)}{\eta^2a^2}\biggl[\frac{(2\alpha-1)r(r+2M)-6M\eta}{(\alpha+1)r^2}+\sqrt{\frac{(\alpha+1)(r+2M)^2+4M\eta}{(\alpha+1)r^2}}\biggl].
\end{align}
${\cal T}_{\rm W}\ge 0$ is satisfied if and only if either of the following two conditions are satisfied: (i) $0<\eta\le \eta_{\rm W}(r)$, (ii) $\eta> \eta_{\rm W}(r)$ and $W_{1}(\eta,r)\le 0$, where
\begin{align}
\eta_{\rm W}(r):=&\frac{(2\alpha-1)r(r+2M)}{6M},\label{def:eta-W}\\
W_1(\eta,r):=&36M^2\eta^2+4Mr\biggl\{6(1-2\alpha)M-(7\alpha-2)r\biggl\}\eta+3\alpha(\alpha-2)r^2(r+2M)^2.
\end{align}
Since $\eta_{\rm W}(r)<\eta_{\rm N}(r)$ holds, the WEC is equivalent to
\begin{align}
0<\eta\le \eta_{\rm W}(r).
\end{align}
For $0<\alpha \le 1/2$, the WEC is violated everywhere because $\eta_{\rm W}(r)$ is non-positive.

Now let us check the DEC for $\alpha>1/2$.
A condition $G^{(0)(0)}- G^{(1)(1)}\ge 0$ is equivalent to $0<\eta\le \eta_{\rm W}(r)$.
Then we compute
\begin{align}
{\cal T}_{\rm D}:=&(G^{(0)(0)}-G^{(1)(1)}-2G^{(2)(2)})+\sqrt{(G^{(0)(0)}+G^{(1)(1)})^2-4(G^{(0)(1)})^2} \nonumber \\
=&\frac{2\alpha(\alpha+1)}{\eta^2a^2}\biggl[\frac{3(\alpha-1)r(r+2M)-6M\eta}{(\alpha+1)r^2}+\sqrt{\frac{(\alpha+1)(r+2M)^2+4M\eta}{(\alpha+1)r^2}}\biggl].
\end{align}
${\cal T}_{\rm D}\ge 0$ is equivalent to $0<\eta\le \eta_{\rm D}(r)$ or $\eta> \eta_{\rm D}(r)$ with $D_1(\eta,r)\le 0$, where
\begin{align}
\eta_{\rm D}(r):=&\frac{(\alpha-1)r(r+2M)}{2M},\\
D_1(\eta,r):=&9M^2\eta^2 - 2Mr\biggl\{9(\alpha - 1)M + (5\alpha - 4)r\biggl\}\eta \nonumber \\
& +(2\alpha-1)(\alpha-2)r^2(r + 2M)^2.
\end{align}
$\eta_{\rm D}(r)\ge \eta_{\rm W}(r)$ is shown to hold for $\alpha\ge 2$ by the following expression:
\begin{align}
\eta_{\rm W}-\eta_{\rm D}=\frac{(2-\alpha)r(r+2M)}{6M}.
\end{align}
Therefore, for $\alpha\ge 2$, the DEC is equivalent to $0<\eta\le \eta_{\rm W}(r)$.

On the other hand, $\eta_{\rm D}(r)< \eta_{\rm W}(r)$ holds for $1/2<\alpha< 2$.
$D_1(\eta,r)=0$ is solved to give $\eta=\eta_{{\rm D}(\pm)} (r)$, where
\begin{align}
\eta_{{\rm D}(\pm)}(r):=&\frac{r}{9M}\biggl\{9(\alpha-1)M + (5\alpha-4)r \nonumber\\
&\pm \sqrt{(\alpha + 1)[(7\alpha-2) r^2 + 18\alpha M r +9(\alpha +1)M^2]}\biggl\},\label{def:eta+-}
\end{align}
which are real and $\eta_{{\rm D}(+)}(r)>\eta_{\rm D}(r)$ is satisfied.
Since the inequality $D_1(0,r)< 0$ shows $\eta_{{\rm D}(-)}(r)<0$, the DEC is equivalent to $0<\eta\le \min\{\eta_{\rm W}(r),\eta_{{\rm D}(+)}(r)\}$ for $1/2<\alpha< 2$.

Lastly, to check the SEC for $\alpha>0$, we compute
\begin{align}
{\cal T}_{\rm S}:=&2G^{(2)(2)}+\sqrt{(G^{(0)(0)}+G^{(1)(1)})^2-4(G^{(0)(1)})^2} \nonumber \\
=&\frac{2\alpha(\alpha+1)}{\eta^2a^2}\biggl[\frac{(2-\alpha)(r+2M)}{(\alpha+1)r}+\sqrt{\frac{(\alpha+1)(r+2M)^2+4M\eta}{(\alpha+1)r^2}}\biggl].
\end{align}
For $0<\alpha\le 2$, the SEC is equivalent to $0<\eta\le \eta_{\rm N}(r)$ because the first term in the large bracket is non-negative.
For $\alpha> 2$, ${\cal T}_{\rm S}\ge 0$ is equivalent to 
\begin{align}
&\eta\ge \frac{3(1-2\alpha)(r+2M)^2}{4(1+\alpha)M}.
\end{align}
Since the right-hand side is negative, the SEC is equivalent also to $0<\eta\le \eta_{\rm N}(r)$ for $\alpha> 2$.
\begin{table}[htb]
\begin{center}
\caption{\label{table:EC-SD-total} Regions where the energy-momentum tensor $T_{\mu\nu}(=G_{\mu\nu}/8\pi)$ respects the energy conditions in the domain $\eta>0$ in the Sultana-Dyer spacetime (\ref{metricSD}) with $M>0$ and $\alpha>0$.}
\begin{tabular}{|c|c|c|c|c|}
\hline
& $0<\alpha\le 1/2$ & $1/2<\alpha< 2$ & $\alpha \ge 2$ \\ \hline\hline
NEC & $0<\eta\le \eta_{\rm N}(r)$ & $0<\eta\le \eta_{\rm N}(r)$ & $0<\eta\le \eta_{\rm N}(r)$ \\ \hline
WEC & nowhere & $0<\eta\le \eta_{\rm W}(r)$ & $0<\eta\le \eta_{\rm W}(r)$ \\ \hline
DEC & nowhere & $0<\eta\le \min\{\eta_{\rm W}(r),\eta_{{\rm D}(+)}(r)\}$ & $0<\eta\le \eta_{\rm W}(r)$ \\ \hline
SEC & $0<\eta\le \eta_{\rm N}(r)$ & $0<\eta\le \eta_{\rm N}(r)$ & $0<\eta\le \eta_{\rm N}(r)$ \\
\hline
\end{tabular}
\end{center}
\end{table}

We have clarified the energy conditions for $T_{\mu\nu}$ in the Sultana-Dyer spacetime (\ref{metricSD}).
The results are summarized in Table~\ref{table:EC-SD-total}, which shows that all standard energy conditions are violated in the region with a finite $r$ for sufficiently large $\eta$.
In particular, Eq.~(\ref{NEC-SD}) gives $\eta_{\rm N}(2M)=28(\alpha+1)M/9$ on the event horizon $r=2M$.
Since at least the NEC should be respected on and outside the event horizon, the Sultana-Dyer class of spacetimes may be a proper model of a cosmological black hole in general relativity only in the early time $0<\eta<28(\alpha+1)M/9$.
Next we will consider decompositions of $T_{\mu\nu}$ into physically motivated matter fields.

\subsubsection{Sultana-Dyer-type decomposition}

In~\cite{Sultana:2005tp}, Sultana and Dyer demonstrated in the case of $\alpha=2$ that the Einstein tensor (\ref{Einstein-SD1})--(\ref{Einstein-SD3}) is compatible with a combination of a dust fluid and a null dust.
This decomposition of $T_{\mu\nu}$ can be generalized to a combination of a perfect fluid and a null dust for arbitrary $\alpha>0$ as
\begin{align}
T^{\mu\nu} =& T^{\mu\nu}_{\rm A} + T^{\mu\nu}_{\rm B}, \label{emt}\\
T^{\mu\nu}_{\rm A} :=& (\rho_{A} + p_{A})u^{\mu}u^{\nu} + p_{A}g^{\mu\nu},\\
T^{\mu\nu}_{\rm B} :=& \rho_{B}k^{\mu}k^{\nu},
\end{align}
where
\begin{align}
u^{\mu} &= \biggl( \frac{r^2 + M(2r -3b)}{ra \sqrt{r^2 + 2M(r - 3b)}}, \frac{M(3b -2r)}{ra \sqrt{r^2 + 2M (r - 3b)}}, 0,0 \biggl),\label{u-SD}\\
8\pi\rho_{\rm A} &= 3 \biggl(1+\frac{2M}{r}\biggl)\frac{\dot a^2}{a^4}- \frac{12M}{r^2}\frac{\dot a}{a^3},\qquad 8\pi p_{\rm A} = \biggl(1+\frac{2M}{r}\biggl)\biggl(-2\frac{{\ddot a}}{a^3} +\frac{{\dot a}^2}{a^4}\biggl)\label{P}
\end{align}
and
\begin{align}
k^{\mu} &= \biggl( \frac{\sqrt{r^2 + 2M(r - 3b)}}{ra}, -\frac{\sqrt{r^2 + 2M(r - 3b)}}{ra}, 0,0 \biggl),\label{k-SD}\\
8\pi\rho_{\rm B} &= \frac{2M \dot{a} \{4r^2 + 3M(2r - 3b)\}}{r^2 a^3\{r^2 + 2M(r - 3b)\}}. \label{rho_{B}}
\end{align}
Here $k_\mu k^\mu=0$ and $u_\mu u^\mu=-1$ hold and $b(\eta)$ is defined by
\begin{align}
b(\eta) := \frac{ a \dot{a}}{2\dot{a}^2 - a \ddot{a}}= \frac{\eta}{\alpha+1}.\label{def-b}
\end{align}
In the asymptotically FLRW region $r\to \infty$, $T^{\mu\nu}_{\rm A}$ becomes homogeneous and $T^{\mu\nu}_{\rm B}\to 0$ is realized.
In the Sultana-Dyer case ($\alpha=2$), in particular, $T^{\mu\nu}_{\rm A}$ becomes a dust fluid ($p_{\rm A}=0$).

However, the problem of this Sultana-Dyer-type decomposition is that $u^\mu$ and $k^\mu$ become complex in the region where the following inequality is satisfied
\begin{align}
b(\eta) > \frac{r(r + 2M)}{6M}~~\to~~ \eta >\frac{(1+\alpha)r(r + 2M)}{6M}=:\eta_{\rm max}. \label{b_max}
\end{align}
Therefore, the decomposition (\ref{emt}) is justified to describe a cosmological black hole, corresponding to the region I$+$II in Fig.~\ref{Sultana-Dyer}, only in the domain $\eta\le \eta_{\rm max}$.

\subsubsection{A global decomposition}

Actually, the Einstein tensor (\ref{Einstein-SD1})--(\ref{Einstein-SD3}) is also compatible with the following energy-momentum tensor that is a combination of a perfect fluid and a type-II null fluid:
\begin{align}
T^{\mu\nu}=&T_{\mathcal{A}}^{\mu\nu}+T_{\mathcal{B}}^{\mu\nu},\label{emt-SD2}\\
T_{\mathcal{A}}^{\mu\nu}:=&\rho_{F} {\tilde u}^\mu {\tilde u}^\nu+p_{F}(g^{\mu\nu}+{\tilde u}^\mu {\tilde u}^\nu),\\
T_{\mathcal{B}}^{\mu\nu}=&\Omega l^\mu l^\nu+(\mu+P)(l^\mu n^\nu+l^\nu n^\mu)+Pg^{\mu\nu},
\end{align}
where $\rho_{F}$, $p_{F}$, $\mu$, $P$, and $\Omega$ are given by Eqs.~(\ref{rho-p-F-def})--(\ref{emt-SD2-last}) and 
\begin{align}
&{\tilde u}_{\mu}\D x^\mu=-a\biggl[\biggl(1-\frac{M}{r}\biggl)\D \eta-\frac{M}{r}\D r\biggl],\\
&l_\mu\D x^\mu=-\frac{1}{\sqrt{2}}a(\D \eta+\D r),\\
&n_{\mu}\D x^\mu =-\frac{1}{\sqrt{2}}a\biggl[\biggl(1-\frac{2M}{r}\biggl)\D \eta-\biggl(1+\frac{2M}{r}\biggl)\D r\biggl],
\end{align}
which satisfy ${\tilde u}_\mu {\tilde u}^\mu=-1$, $l_\mu l^\mu=n_\mu n^\mu=0$, and $l_\mu n^\mu=-1$.
The decomposition (\ref{emt-SD2}) is valid in the entire Sultana-Dyer spacetime, where the perfect fluid $T_{\mathcal{A}}^{\mu\nu}$ is homogeneous, while the type II null fluid $T_{\mathcal{B}}^{\mu\nu}$ is inhomogeneous.

Adopting the orthonormal basis one-forms (\ref{E0-SD})--(\ref{E2-SD}), we obtain
\begin{align}
T_{\mathcal{A}}^{(a)(b)}=T_{\mathcal{A}}^{\mu\nu}E_\mu^{(a)}E_\nu^{(b)}=\mbox{diag}(\rho_{F},p_{F},p_{F},p_{F})
\end{align}
and
\begin{align}
\label{T-typeII}
T_{\mathcal{B}}^{(a)(b)}=T_{\mathcal{B}}^{\mu\nu}E_\mu^{(a)}E_\nu^{(b)}=
\begin{pmatrix}
\mu+\Omega/2 &\Omega/2&0&0\\
\Omega/2&-\mu+\Omega/2&0&0\\
0&0&P&0 \\
0&0&0&P
\end{pmatrix}
.
\end{align}
Thus, by Eqs.~(\ref{NEC-I})--(\ref{SEC-I}), the homogeneous perfect fluid $T_{\mathcal{A}}^{\mu\nu}$ satisfies the standard energy conditions according to Table~\ref{table:EC-FLRW}.
On the other hand, Eqs.~(\ref{mu-P-def}) and (\ref{emt-SD2-last}) with $a(\eta)=a_0|\eta|^\alpha$ give
\begin{align}
&8\pi\mu=\frac{2\alpha M[(2\alpha-1)r-3\eta]}{r^2\eta^2a^2},\qquad 8\pi P=-\frac{2\alpha(\alpha-2) M}{r\eta^2a^2},\\
&4\pi \Omega=\frac{2\alpha M[(\alpha+1)(r+M)+\eta]}{r^2\eta^2a^2},
\end{align}
so that
\begin{align}
&8\pi(\mu+P)=\frac{2\alpha M[(\alpha+1)r-3\eta]}{r^2\eta^2a^2},\\
&8\pi(\mu-P)=\frac{6\alpha M[(\alpha-1)r-\eta]}{r^2\eta^2a^2}.
\end{align}
With the above expressions, we can check whether $T_{\mathcal{B}}^{\mu\nu}$ respects the energy conditions or not according to Eqs.~(\ref{NEC-II})--(\ref{SEC-II}).
The results are summarized in Table~\ref{table:EC-SD}.
\begin{table}[htb]
\begin{center}
\caption{\label{table:EC-SD} Regions where the type-II null fluid $T_{\mathcal{B}}^{\mu\nu}$ respects the energy conditions in the domain $\eta>0$ in the Sultana-Dyer spacetime (\ref{metricSD}) with $M>0$ and $\alpha>0$.}
\scalebox{0.9}{
\begin{tabular}{|c|c|c|c|c|}
\hline
& $0<\alpha\le 1/2$ & $1/2<\alpha\le 1$ & $1<\alpha \le 2$ & $\alpha > 2$ \\ \hline\hline
NEC & {$0<\eta\le (\alpha+1)r/3$} & {$0<\eta\le (\alpha+1)r/3$} & {$0<\eta\le (\alpha+1)r/3$} & {$0<\eta\le (\alpha+1)r/3$} \\ \hline
WEC & {nowhere} & {$0<\eta\le (2\alpha-1)r/3$} & {$0<\eta\le (2\alpha-1)r/3$} & {$0<\eta\le (\alpha+1)r/3$} \\ \hline
DEC & {nowhere} & {nowhere} & {$0<\eta\le (\alpha-1)r$} & {$0<\eta\le (\alpha+1)r/3$} \\ \hline
SEC & {$0<\eta\le (\alpha+1)r/3$} & {$0<\eta\le (\alpha+1)r/3$} & {$0<\eta\le (\alpha+1)r/3$} & {nowhere} \\
\hline
\end{tabular}
}
\end{center}
\end{table}

\subsection{Properties as a cosmological black hole}

Now we study properties of the cosmological black-hole spacetime with the metric (\ref{metricSD}) with $a(\eta) = a_{0}\eta^{\alpha}$ ($\alpha>0$) which corresponds to the region I$+$II in Fig.~\ref{Sultana-Dyer} ($\eta>0$).

\subsubsection{Trapping horizon}

Here we identify the locations of trapping horizons and their types.
Consider a future-directed outgoing and ingoing radial null vectors $k^\mu$ and $l^\mu$ given by 
\begin{align}
k^\mu\frac{\partial}{\partial x^\mu}=&\frac{1}{\sqrt{2}a}\biggl\{\biggl(1+\frac{2M}{r}\biggl)\frac{\partial}{\partial \eta}+\biggl(1-\frac{2M}{r}\biggl)\frac{\partial}{\partial r}\biggl\},\\
l^\mu\frac{\partial}{\partial x^\mu}=&\frac{1}{\sqrt{2}a}\biggl(\frac{\partial}{\partial \eta}-\frac{\partial}{\partial r}\biggl),
\end{align}
respectively, which satisfy $k_\mu k^\mu=l_\mu l^\mu=0$ and $k_\mu l^\mu=-1$.
With the areal radius $R=ar$, the expansions (\ref{exp+}) and (\ref{exp-}) are computed to give
\begin{align}
\theta_+=&\frac{\sqrt{2}}{ar^2}\biggl\{(r-2M)+\frac{\alpha r(r+2M)}{\eta}\biggl\},\\
\theta_-=&\frac{\sqrt{2}}{ar}\biggl(\frac{\alpha r}{\eta}-1\biggl),
\end{align}
which show that trapping horizons associated with $k^\mu$ and $l^\mu$ are respectively given by $\eta=\eta_+(r)$ and $\eta=\eta_-(r)$, where
\begin{align}
\eta_+(r):=-\frac{\alpha r(r+2M)}{r-2M},\qquad \eta_-(r):=\alpha r.\label{eta-pm}
\end{align}
Note that $\eta_+(r)>0$ holds only in the domain $0<r<2M$.

The line element along $\eta=\eta_+(r)$ on $(M^2,g_{AB})$ is given by 
\begin{align}
\D s^2|_{\eta=\eta_+}=&\frac{a(\eta_+)^2}{r x^3}\left\{2\alpha r^2 +2 (2\alpha+1)xr+(1+\alpha)x^2\right\} \nonumber \\
&\times \left\{2\alpha r^2 + 4\alpha xr + (1+\alpha)x^2\right\}\D r^2,
\end{align}
where $x:=2M-r$.
Since the trapping horizon $\eta=\eta_+(r)(>0)$ exists only in the domain $0<r<2M$, $\D s^2|_{\eta=\eta_+}>0$ is satisfied and therefore $\eta=\eta_+(r)$ is spacelike.
On the other hand, the line element along $\eta=\eta_-(r)$ on $(M^2,g_{AB})$ is given by 
\begin{align}
\D s^2|_{\eta=\eta_-}=&\frac{(1+\alpha)a(\eta_-)^2[(1-\alpha)r+2M(1+\alpha)]}{r}\D r^2.
\end{align}
For $0<\alpha\le 1$, $\D s^2|_{\eta=\eta_-}>0$ is satisfied, so that $\eta=\eta_-(r)$ is spacelike.
For $\alpha>1$, $\eta=\eta_-(r)$ is timelike, null, and spacelike in the domains $r>r_{\rm d(SD)}$, $r=r_{\rm d(SD)}$, and $0<r<r_{\rm d(SD)}$, respectively, where 
\begin{align}
r_{\rm d(SD)}:=\frac{2M(\alpha+1)}{\alpha-1}(>2M).
\end{align}
The results are summarized in Table~\ref{table:signature-TH-SD}.
\begin{table}[htb]
\begin{center}
\caption{\label{table:signature-TH-SD} Signature of the trapping horizon $\eta=\eta_-(r)$ in the Sultana-Dyer spacetime.}
\begin{tabular}{|c|c|c|c|c|}
\hline
& timelike & null & spacelike \\ \hline\hline
$0<\alpha\le 1$ & n.a. & n.a. & any $r(>0)$ \\ \hline
$\alpha>1$ & $r>r_{\rm d(SD)}$ & $r=r_{\rm d(SD)}$ & $0<r<r_{\rm d(SD)}$ \\ 
\hline
\end{tabular}
\end{center}
\end{table}

\begin{figure}[htbp]
\begin{center}
\includegraphics[width=0.7\linewidth]{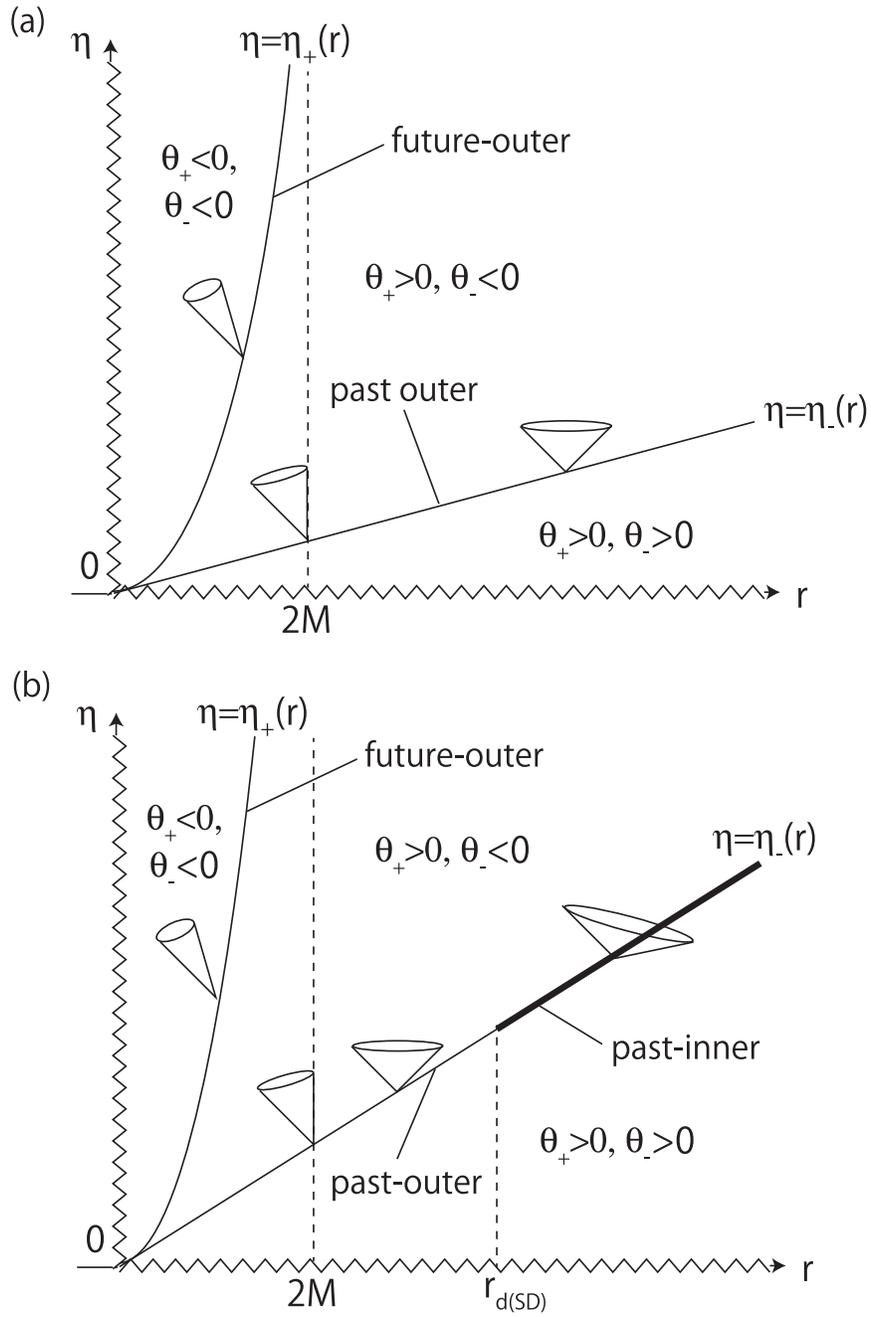}
\caption{\label{Fig:SD-r-eta-total} $(\eta,r)$ planes for the Sultana-Dyer class of spacetimes with $a(\eta)=a_0\eta^{\alpha}$ for (a) $0<\alpha\le 1$ and 
(b) $\alpha>1$. 
Several future light-cones are put to clarify the signature of the trapping horizons $\eta=\eta_\pm(r)$. A thick portion of $\eta=\eta_{-}(r)$ is a past inner trapping horizon.}
\end{center}
\end{figure}

In the $(\eta,r)$-plane shown in Fig.~\ref{Fig:SD-r-eta-total}, one recognizes that ${\cal L}_-\theta_+<0$ is satisfied at $\eta=\eta_+(r)$ defined by $\theta_+=0$ and therefore $\eta=\eta_+(r)$ is a future outer trapping horizon for any $\alpha(>0)$.
On the other hand, at $\eta=\eta_-(r)$ defined by $\theta_-=0$, ${\cal L}_+\theta_-<(>)0$ is satisfied when $\eta=\eta_-(r)$ is spacelike (timelike) and then the trapping horizon $\eta=\eta_-(r)$ is past outer (past inner).
Therefore, $\eta=\eta_-(r)$ is past outer for $0<\alpha\le 1$.
For $\alpha>1$, $\eta=\eta_-(r)$ is past outer, past degenerate, and past inner in the domain $0<r<r_{\rm d(SD)}$, $r=r_{\rm d(SD)}$, and $r>r_{\rm d(SD)}$, respectively.

\begin{figure}[htbp]
\begin{center}
\includegraphics[width=0.5\linewidth]{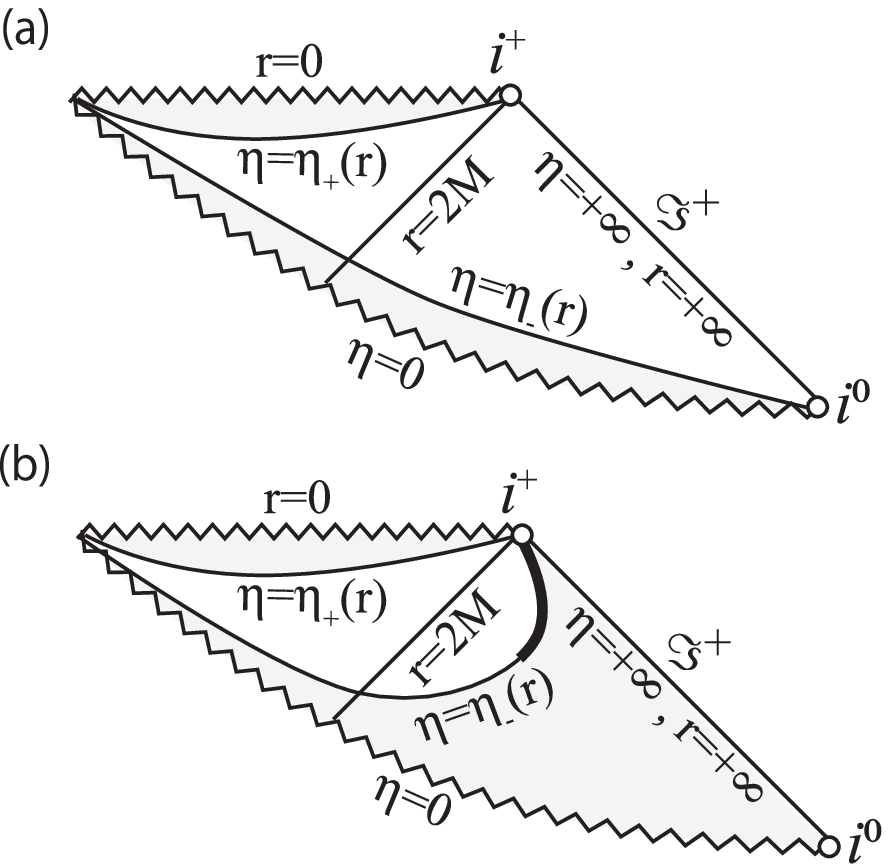}
\caption{\label{Fig:Sultana-Dyer-TH} Penrose diagrams of the Sultana-Dyer black-hole spacetime with $a(\eta)=a_0\eta^{\alpha}$ for (a) $0<\alpha\le 1$ and 
(b) $\alpha>1$. 
Shaded regions are trapped regions. A thick timelike portion of $\eta=\eta_-(r)$ is a past inner trapping horizon. }
\end{center}
\end{figure}

Collecting all the information obtained, one can draw the Penrose diagrams of the Sultana-Dyer black hole as in Fig.~\ref{Fig:Sultana-Dyer-TH}.
We confirm the asymptotic behaviors of the trapping horizons $\eta=\eta_\pm(r)$ as follows.
Using Eqs.~(\ref{uv-def}), (\ref{r*-def}), (\ref{UV-def}), (\ref{UV-r}), and (\ref{def-eta-KS}), one can write the compactified Kruskal-Szekeres coordinates (\ref{barUV-def}) in terms of $(\eta,r)$ as
\begin{align}
{\bar U}&= \arctan \left[ \biggl(1 - \frac{r}{2 M}\biggl) e^{(r - \eta)/(4M)}\right],\label{barU-SD}\\
{\bar V} &= \arctan(e^{(r + \eta)/(4M)}).\label{barV-SD}
\end{align}
Substituting $\eta=\eta_+(r)$ defined by Eq.~(\ref{eta-pm}), we obtain
\begin{align}
{\bar U}&= \arctan \biggl[ \biggl(1 - \frac{r}{2M} \biggl) \exp \biggl\{ \frac{r}{4M} \biggl(1 + \frac{\alpha(r + 2M)}{r - 2M} \biggl)\biggl\} \biggl],\\
{\bar V} &= \arctan \biggl[ \exp \biggl\{ \frac{r}{4M} \biggl(1 - \frac{\alpha(r + 2M)}{r - 2M} \biggl) \biggl\} \biggl],
\end{align}
which converge to $({\bar U},{\bar V})\to (0,\pi/2)$ as $r\to 2{M^-}$. (See Fig.~\ref{Sultana-Dyer}.)
On the other hand, substituting $\eta=\eta_-(r)$ defined by Eq.~(\ref{eta-pm}) into Eqs.~(\ref{barU-SD}) and (\ref{barV-SD}), we obtain
\begin{align}
{\bar U}&= \arctan \left[ \biggl(1 - \frac{r}{2 M}\biggl) e^{(1 - \alpha)r/(4M)} \right],\\
{\bar V} &= \arctan(e^{(1 + \alpha)r/(4M)}).
\end{align}
The asymptotic behaviors as $r \to \infty$ are $({\bar U},{\bar V}) \to(-\pi/2,\pi/2)$ for $0 < \alpha \leq 1$ and $({\bar U},{\bar V}) \to(0,\pi/2)$ for $\alpha>1$.
Hence, $\eta=\eta_-(r)$ approaches the spacelike infinity $i^0$ in Fig.~\ref{Sultana-Dyer} for $0 < \alpha \leq 1$ and the future timelike infinity $i^+$ for $\alpha>1$.

\subsubsection{Misner-Sharp mass}

The Misner-Sharp mass (\ref{MS-mass}) for the Sultana-Dyer spacetime is computed to give
\begin{align}
m_{\rm MS}(\eta,r)&= Ma - 2Mr\dot{a} + \frac{r^2(r + 2M)\dot{a}^2}{2a} \notag \\
&= Ma \biggl(1 - \frac{2\alpha r}{\eta} + \frac{\alpha^2 r^2}{\eta^2} + \frac{\alpha^2 r^3}{2M\eta^2}\biggl).\label{MSmass}
\end{align}
At the central singularity $r=0$, the Misner-Sharp mass is positive $m_{\rm MS}(\eta,0)=a(\eta)M(>0)$.
It is also positive on the event horizon ($r=2M$) in the domain $0<\eta<\infty$:
\begin{align}
m_{\rm MS}(\eta,2M)= Ma\biggl(1 - \frac{4\alpha M}{\eta} + \frac{8\alpha^2 M^2}{\eta^2}\biggl)>0.
\end{align}

Differentiating Eq.~(\ref{MSmass}) with respect to $r$, we obtain
\begin{align}
\partial_r{m_{\rm MS}}= - 2M\dot{a} + \frac{r(3r + 4M)\dot{a}^2}{2a}=\frac{{\dot a}}{2}\biggl\{ - 4M + \frac{\alpha r(3r + 4M)}{\eta}\biggl\},
\end{align}
which shows $\partial_r{m_{\rm MS}}<0$ in the region of
\begin{align}
\eta>\frac{\alpha r(3r + 4M)}{4M}.
\end{align}
Figures~\ref{Fig:SD-r-eta-total} and \ref{Fig:Sultana-Dyer-TH} show that the event horizon $r=2M$ is in an untrapped region for $\eta>\eta_-(2M)(=2\alpha M)$.
Since the contraposition of Proposition~\ref{th:monotonicity} asserts that the DEC is violated in an untrapped region with $\partial_r{m_{\rm MS}}<0$, the DEC is violated on the event horizon $r=2M$ in the late time $\eta>5\alpha M$.

This result is consistent with Table~\ref{table:EC-SD-total} for $\alpha>1/2$.
Equations~(\ref{def:eta-W}) and (\ref{def:eta+-}) give
\begin{align}
\eta_{\rm W}(2M)=&\frac{4(2\alpha-1)M}{3},\\
\eta_{{\rm D}+}(2M)=&\frac{2}{9}M\biggl\{19\alpha-17 +\sqrt{(\alpha + 1)(73\alpha +1)}\biggl\},
\end{align}
which satisfy $\eta_{{\rm D}+}(2M)> \eta_{\rm W}(2M)$ for $\alpha> 1/2$.
Therefore, the DEC is violated on the event horizon in the domain $\eta> \eta_{\rm W}(2M)$ for $\alpha>1/2$.
Since $5\alpha M>\eta_{\rm W}(2M)$ is satisfied for $\alpha>0$, the DEC is certainly violated on the event horizon $r=2M$ in the late time $\eta>5\alpha M$.

\subsubsection{Temperature of a cosmological black hole}

Since the conformal Killing vector $\xi^\mu=(1,0,0,0)$ in the Sultana-Dyer spacetime satisfies $\xi_\mu \xi^\mu=-a^2(1-2M/r)$, the event horizon $r=2M$ is a conformal Killing horizon.
Its temperature (\ref{T-CK}) is computed to give
\begin{align}
T_{\rm JKSD}=\frac{1}{8\pi M},
\end{align}
which is the same as the temperature of a hypersurface-orthogonal Killing vector in the Schwarzschild black-hole spacetime.

Next, we derive the temperature of the future outer trapping horizon $\eta=\eta_+(r)$ given by Eq.~(\ref{eta-pm}).
In the spacetime (\ref{metricSD}), the Kodama vector (\ref{kodamavector}) is given by 
\begin{align}
K^\mu\frac{\partial}{\partial x^\mu}=\frac{1}{a}\frac{\partial}{\partial \eta}-\frac{r{\dot a}}{a^2}\frac{\partial}{\partial r},
\end{align}
which reduces to a hypersurface-orthogonal Killing vector $\xi^\mu=(1,0,0,0)$ in the static case ($a(\eta)\equiv 1$).
The temperature (\ref{T-TH}) of $\eta=\eta_+(r)$ is computed to give
\begin{align}
T_{\rm TH}=&\frac{1}{2\pi a(\eta_+(r))}\times\frac{2\alpha r(4M-r) +(1+\alpha)(2M-r)^2}{2\alpha r^2(r+2M)} \nonumber \\
=&\frac{1}{8\pi M}\times\frac{h(x)}{a_0(2M)^{\alpha}},
\end{align}
where $h(x)$ is a dimensionless function of $x:=r/(2M)$ defined by 
\begin{align}
h(x):=\frac{(1-x)^\alpha[2\alpha x(2-x) +(1+\alpha)(1-x)^2]}{\alpha^{\alpha+1}x^{\alpha+2}(x+1)^{\alpha+1}}.
\end{align}
$T_{\rm TH}$ is a positive function of $r$ in the domain $0<r<2M$ and it diverges as $(\eta,r)\to (0,0)$ and converges to zero as $(\eta,r)\to (\infty,2M)$.

In this section, we have fully investigated the Sultana-Dyer class of spacetimes (\ref{metricSD}) and the corresponding matter field with $a(\eta)=a_0\eta^{\alpha}~(\alpha>0)$.
In fact, this class of spacetimes can be generalized further to be non-conformally Schwarzschild as shown in Appendix~\ref{app:Husain}.
The generalized solution, which is conformally related to the Husain solution~\cite{Husain:1995bf}, is also a candidate of a more general cosmological black-hole spacetime.

\section{Culetu class}
\label{sec:Culetu}

In~\cite{Culetu:2012ih}, Culetu studied a conformally Schwarzschild spacetime with $M>0$, constructed with the Painlev\'{e}-Gullstrand coordinates (\ref{Sch-P}) such as
\begin{align}
\D s^2=&a(\tau)^2\biggl[-\biggl(1-\frac{2M}{r}\biggl)\D \tau^2+2\sqrt{\frac{2M}{r}}\D \tau \D r+\D r^2+r^2\D \Omega^2\biggl],\label{PG-c}
\end{align}
which is asymptotic to the flat FLRW spacetime as $r\to \infty$.
Non-zero components of the inverse metric are given by 
\begin{align}
\begin{aligned}
&g^{\tau\tau}=-a^{-2},\qquad g^{rr}=\frac{1}{a^2}\biggl(1-\frac{2M}{r}\biggl), \\
&g^{\tau r}(=g^{r\tau})=\frac{1}{a^2}\sqrt{\frac{2M}{r}},\qquad g^{\theta\theta}=g^{\phi\phi}\sin^2\theta=\frac{1}{a^2r^2}.
\end{aligned}
\end{align}
Culetu found that $r=2M$ is not a curvature singularity and pointed out that the spacetime can be a model of a cosmological black hole.
In addition, in spite that there is a non-zero off-diagonal component ${G^1}_0$ of the Einstein tensor, he showed that the corresponding matter field may be interpreted as an anisotropic fluid.
In the Culetu spacetime, $\tau$ is a timelike coordinate everywhere and we define the future direction by increasing $\tau$.
In this section, we will focus on the case with $a(\tau) = a_{0}|\tau|^{\alpha}$ where $a_{0}$ and $\alpha$ are positive constants.

It is noted that, by a coordinate transformation (\ref{r-Lemaitre}), the Culetu spacetime can be expressed in the Lema\^{\i}tre coordinates as
\begin{eqnarray}
\D s^2=a(\tau)^2\biggl[-\D \tau^2+(2M)^{2/3}\biggl\{\frac{\D \chi^2}{\left[\frac32(\chi-\tau)\right]^{2/3}}+\left[\frac32(\chi-\tau)\right]^{4/3}\D \Omega^2\biggl\}\biggl]. \label{Lemaitre-c}
\end{eqnarray}
Although the FLRW limit $M\to 0$ is singular, the spacetime is asymptotic to the flat FLRW spacetime as $\chi\to \infty$, which is confirmed with a radial coordinate $r=(2M)^{1/3}(3\chi/2)^{2/3}$.

\subsection{Global structure}

Under our assumptions, the spacetime with the metric (\ref{PG-c}) is analytic except at $\tau=0$ and $r=0$, while the spacetime with the metric (\ref{Lemaitre-c}) is analytic except at $\tau=0$ and $\tau=\chi$.
The Ricci scalar ${\cal R}$ is given by 
\begin{align}
{\cal R} =& \frac{3}{a^2}\biggl\{2\frac{{\ddot a}}{a}-\frac{3}{r}\sqrt{\frac{2M}{r}}\frac{{\dot a}}{a}\biggl\}= \frac{3\alpha}{a_0^2|\tau|^{2\alpha}}\biggl\{\frac{2(\alpha-1)}{\tau^2}-\frac{3}{r}\sqrt{\frac{2M}{r}}\frac{1}{\tau}\biggl\} \label{Ricci-s-PG}
\end{align}
in the Painlev\'{e}-Gullstrand coordinates (\ref{PG-c}) and 
\begin{align}
{\cal R} =\frac{6}{(\chi-\tau)a^2}\biggl\{(\chi-\tau)\frac{{\ddot a}}{a}-\frac{{\dot a}}{a}\biggl\}=\frac{6\alpha[(\alpha-1)\chi-\alpha\tau]}{(\chi-\tau)a_0^2|\tau|^{2(\alpha+1)}}
\end{align}
in the Lema\^{\i}tre coordinates (\ref{Lemaitre-c}).
Since ${\cal R}$ blows up, $\tau=0$, $r=0$, and $\tau=\chi$ are curvature singularities.
Hence, in the Lema\^{\i}tre coordinates (\ref{Lemaitre-c}), the domains of $\tau$ and $\chi$ including the asymptotically FLRW region are ${\tau}\in(0,\infty)$ and $\chi>\tau$.

Because the spacetime with the metric (\ref{PG-c}) admits a hypersurface-orthogonal conformal Killing vector $\xi^\mu=(1,0,0,0)$ satisfying ${\cal L}_\xi g_{\mu\nu}=2({\dot a}/a)g_{\mu\nu}$, there is a conserved quantity $C:=-\xi_\mu {\bar k}^\mu$ along a null geodesic with its tangent vector ${\bar k}^\mu$.
A future-directed radial null geodesic $\gamma$ is described by $x^\mu=(\tau(\lambda), r(\lambda),0,0)$, where $\lambda$ is an affine parameter along $\gamma$.
Then, we have ${\bar k}^\mu=(\D \tau/\D\lambda, \D r/\D\lambda,0,0)$ and 
\begin{equation}
C=a(\tau)^2\biggl[\biggl(1-\frac{2M}{r}\biggl){\bar k}^0-\sqrt{\frac{2M}{r}}{\bar k}^1\biggl].\label{C-PG}
\end{equation}
By the null condition $\D s^2=0$ for $\gamma$, we obtain
\begin{align}
{\bar k}^1=\biggl(1-\sqrt{\frac{2M}{r}}\biggl){\bar k}^0 \label{k-out}
\end{align}
for outgoing $\gamma$ and 
\begin{align}
{\bar k}^1=-\biggl(1+\sqrt{\frac{2M}{r}}\biggl){\bar k}^0 \label{k-in}
\end{align}
for ingoing $\gamma$.
Equation~(\ref{k-out}) is integrated to give
\begin{align}
\tau-\tau_0=&r+2\sqrt{2Mr}+4M\ln\biggl|{\sqrt{\frac{r}{2M}}-1}\biggl|,\label{k-out-int}
\end{align}
where $\tau_0$ is an integration constant.
Equation~(\ref{k-out-int}) shows $\tau\to -\infty$ as $r\to 2M$ along an outgoing $\gamma$ as seen in Fig.~\ref{Schwarzschild-3coordinates}(a).
Equations~(\ref{C-PG}) and (\ref{k-out}) give
\begin{equation}
\frac{C}{a(\tau)^2}={\bar k}^1.\label{C-int}
\end{equation}
With $a(\tau)=a_0|\tau|^\alpha$ and Eq.~(\ref{k-out-int}), Eq.~(\ref{C-int}) is integrated to give
\begin{align}
\frac{C}{a_0^2}(\lambda-\lambda_0)=\int^r \biggl|\tau_0+{\bar r}+2\sqrt{2M{\bar r}}+4M\ln\biggl|{\sqrt{\frac{{\bar r}}{2M}}-1}\biggl|\biggl|^{2\alpha}\D {\bar r},\label{F-int}
\end{align}
where $\lambda_0$ is an integration constant.
Since the affine parameter $\lambda$ given by Eq.~(\ref{F-int}) diverges as $r\to\infty$ along a future-directed outgoing radial null geodesic $\gamma$, $(\tau,r)\to (\infty,\infty)$ is a future null infinity.
In contrast, one can show that the right-hand side of Eq.~(\ref{F-int}) is finite as $r\to 2M$.
Since $\lambda$ is finite, $(\tau,r)\to (-\infty,2M)$ is an extendable boundary if it is regular as in the Sultana-Dyer class of spacetimes.

On the other hand, Eq~(\ref{k-in}) for ingoing $\gamma$ is integrated to give
\begin{align}
\tau-\tau_0=&-r+2\sqrt{2Mr}-4M\ln\biggl|{\sqrt{\frac{r}{2M}}+1}\biggl|,\label{k-in-int}
\end{align}
which shows $\tau\to -\infty$ as $r\to \infty$, as seen in Fig.~\ref{Schwarzschild-3coordinates}(a).
Equations~(\ref{C-PG}) and (\ref{k-in}) give
\begin{equation}
\frac{C}{a(\tau)^2}=-{\bar k}^1.\label{C-int-in}
\end{equation}
With $a(\tau)=a_0|\tau|^\alpha$ and Eq.~(\ref{k-in-int}), Eq.~(\ref{C-int-in}) is integrated to give
\begin{equation}
\frac{C}{a_0^2}(\lambda-\lambda_0)=-\int^r \biggl|\tau_0-{\bar r}+2\sqrt{2M{\bar r}}-4M\ln\biggl|{\sqrt{\frac{{\bar r}}{2M}}+1}\biggl|\biggl|^{2\alpha}\D {\bar r},
\end{equation}
{which diverges as $r\to \infty$.
Since $|\lambda|$ diverges along a future-directed ingoing radial null geodesic $\gamma$, $(\tau,r)\to (-\infty,\infty)$ is a past null infinity.}

\begin{figure}[htbp]
\begin{center}
\includegraphics[width=0.7\linewidth]{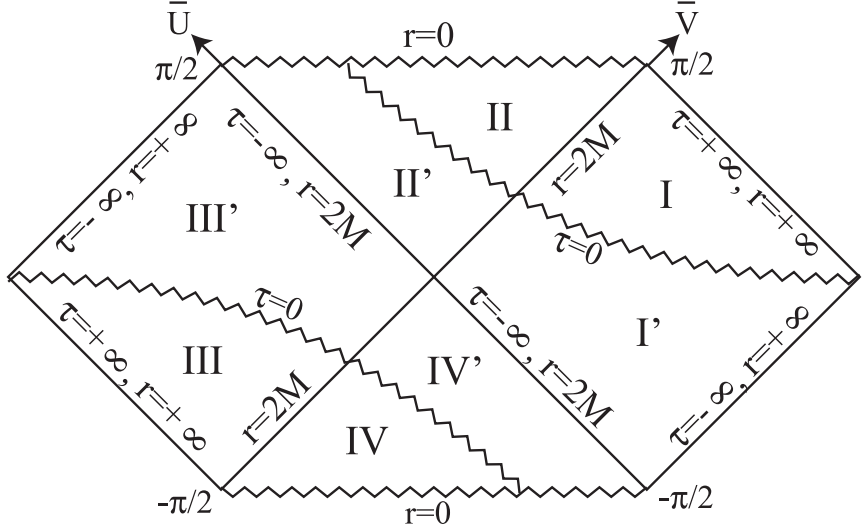}
\caption{\label{Culetu} A Penrose diagram of the maximally extended Culetu spacetime with the metric (\ref{PG-c}) with $a(\tau)=a_0|\tau|^\alpha~(\alpha>0)$. The portion consisting of I and II describes a cosmological black hole. {If the coordinate boundary given by $\tau\to -\infty$ and $r=2M$ is regular, it is extendable.}}
\end{center}
\end{figure}

As a result, the Penrose diagram of the Culetu spacetime (\ref{PG-c}) with $a(\tau)=a_0|\tau|^\alpha~(\alpha>0)$ is drawn as in Fig.~\ref{Culetu}.
It is clear that a maximally extended spacetime given in the domains $\tau>0$ and $r>0$, which corresponds to the portion consisting of I and II, represents an asymptotically flat FLRW cosmological black hole with the event horizon at $r=2M$.
While $r=0$ corresponds to a black-hole singularity, $\tau=0$ corresponds to a big-bang singularity.
Similarly to the Sultana-Dyer class (\ref{metricSD}), the coordinate boundary $(\tau,r)\to (-\infty,2M)$ corresponds to a finite affine parameter along a radial null geodesic.
Therefore, if it is regular, the maximally extended spacetime consisting of I', II', III', and IV' also describes a cosmological black hole.
Even in that case, similar to the Sultana-Dyer class of spacetimes, differentiability of the metric at the event horizon $(\tau,r)\to (-\infty,2M)$ is a non-trivial problem.
Hereafter, we will focus on the cosmological black hole corresponding to I$+$II.

\subsection{Matter fields}

Now let us identify the matter field to give the corresponding energy-momentum tensor $T_{\mu\nu}(=G_{\mu\nu}/8\pi)$ in the cosmological black-hole spacetime in the domains $0<\tau<\infty$ and $0<r<\infty$.
For this purpose, the Lema\^{\i}tre coordinates (\ref{Lemaitre-c}) are useful because $\tau$ is timelike everywhere and the Einstein tensor $G^\mu_{~\nu}$ is diagonal such that
\begin{align}
\begin{aligned}
&G^{\tau}_{~\tau}=-3\frac{{\dot a}^2}{a^4}+\frac{2{\dot a}}{a^3(\chi-\tau)},\quad G^{\chi}_{~\chi}=-2\frac{{\ddot a}}{a^3}+\frac{{\dot a}^2}{a^4}+\frac{8{\dot a}}{3a^3(\chi-\tau)},\\
&G^{\theta}_{~\theta}=G^{\phi}_{~\phi}=-2\frac{{\ddot a}}{a^3}+\frac{{\dot a}^2}{a^4}+\frac{2{\dot a}}{3a^3(\chi-\tau)}.
\end{aligned}
\end{align}
Hence, the corresponding energy-momentum tensor $T_{\mu\nu}$ is of the Hawking-Ellis type I and can be identified as an anisotropic fluid given by 
\begin{align}
&{T}_{\mu\nu}=(\rho+p_{\rm t})u_\mu u_\nu+(p_{\rm r}-p_{\rm t})s_\mu s_\nu +p_{\rm t}g_{\mu\nu},\label{T-fluid-L}\\
&u^\mu\frac{\partial}{\partial x^\mu}=\frac{1}{a}\frac{\partial}{\partial \tau},\qquad s^\mu\frac{\partial}{\partial x^\mu}=\frac{1}{a}\left[\frac{3}{4M}(\chi-\tau)\right]^{1/3}\frac{\partial}{\partial \chi},\\
&8\pi \rho=3\frac{{\dot a}^2}{a^4}-\frac{2{\dot a}}{a^3(\chi-\tau)},\quad 8\pi p_{\rm r}=-2\frac{{\ddot a}}{a^3}+\frac{{\dot a}^2}{a^4}+\frac{8{\dot a}}{3a^3(\chi-\tau)},\\
&8\pi p_{\rm t}=-2\frac{{\ddot a}}{a^3}+\frac{{\dot a}^2}{a^4}+\frac{2{\dot a}}{3a^3(\chi-\tau)}.
\end{align}
This result was obtained by Culetu in the Painlev\'{e}-Gullstrand coordinates (\ref{PG-c})~\cite{Culetu:2012ih}.

By a coordinate transformation (\ref{r-Lemaitre}), the energy-momentum tensor (\ref{T-fluid-L}) is written in the Painlev\'{e}-Gullstrand coordinates (\ref{PG-c}) as
\begin{align}
&u^\mu\frac{\partial}{\partial x^\mu}=\frac{1}{a}\biggl(\frac{\partial}{\partial \tau}-\sqrt{\frac{2M}{r}}\frac{\partial}{\partial r}\biggl),\qquad s^\mu\frac{\partial}{\partial x^\mu}=\frac{1}{a}\frac{\partial}{\partial r},\label{us-Culetu}\\
&8\pi \rho=3\frac{{\dot a}^2}{a^4}-\frac{3{\dot a}}{a^3r}\sqrt{\frac{2M}{r}},\quad 8\pi p_{\rm r}=-2\frac{{\ddot a}}{a^3}+\frac{{\dot a}^2}{a^4}+\frac{4{\dot a}}{a^3r}\sqrt{\frac{2M}{r}},\label{rho-Culetu}\\
&8\pi p_{\rm t}=-2\frac{{\ddot a}}{a^3}+\frac{{\dot a}^2}{a^4}+\frac{{\dot a}}{a^3r}\sqrt{\frac{2M}{r}}.\label{pt-Culetu}
\end{align}
With $a(\tau)=a_0|\tau|^\alpha$, we obtain
\begin{align}
&8\pi a^2(\rho+p_{\rm r})=\frac{2\alpha(\alpha+1)}{\tau^2}+\frac{\alpha}{\tau r}\sqrt{\frac{2M}{r}},\\
&8\pi a^2(\rho+p_{\rm t})=\frac{2\alpha(\alpha+1)}{\tau^2}-\frac{2\alpha}{\tau r}\sqrt{\frac{2M}{r}},\\
&8\pi a^2\rho=\frac{3\alpha^2}{\tau^2}-\frac{3\alpha}{\tau r}\sqrt{\frac{2M}{r}},\\
&8\pi a^2(\rho-p_{\rm r})=\frac{2\alpha(2\alpha-1)}{\tau^2}-\frac{7\alpha}{\tau r}\sqrt{\frac{2M}{r}},\\
&8\pi a^2(\rho-p_{\rm t})=\frac{2\alpha(2\alpha-1)}{\tau^2}-\frac{4\alpha}{\tau r}\sqrt{\frac{2M}{r}},\\
&8\pi a^2(\rho+p_{\rm r}+2p_{\rm t})=\frac{6\alpha}{\tau^2}+\frac{3\alpha}{\tau r}\sqrt{\frac{2M}{r}},
\end{align}
which show $\rho+p_{\rm r}\ge 0$ and $\rho+p_{\rm r}+2p_{\rm t}\ge 0$ in the domain $\tau>0$.
Then, from the following equivalent expressions:
\begin{align}
&\rho+p_{\rm t}\ge 0~~\leftrightarrow~~0<\tau\le (\alpha+1)\sqrt{\frac{r^3}{2M}},\\
&\rho\ge 0~~\leftrightarrow~~0<\tau\le \alpha\sqrt{\frac{r^3}{2M}},\\
&\rho-p_{\rm r}\ge 0~~\leftrightarrow~~0<\tau\le \frac{2(2\alpha-1)}{7}\sqrt{\frac{r^3}{2M}},\\
&\rho-p_{\rm t}\ge 0~~\leftrightarrow~~0<\tau\le \frac{2\alpha-1}{2}\sqrt{\frac{r^3}{2M}},
\end{align}
we can identify the regions where the energy-momentum tensor (\ref{T-fluid-L}) respects the energy conditions in the domain $\tau>0$.
The results are summarized in Table~\ref{table:EC-Culetu}, which shows that all the standard energy conditions are violated in the region with a finite $r$ for sufficiently large $\tau$.
In particular, the NEC is violated on the event horizon $r=2M$ in the late time $\tau> 2(\alpha+1)M$.
This implies that the Culetu spacetime may be a proper model of a cosmological black hole in general relativity only in the early time.
\begin{table}[htb]
\begin{center}
\caption{\label{table:EC-Culetu} Regions where the energy-momentum tensor (\ref{T-fluid-L}) respects the energy conditions in the domain $\tau>0$ in the Culetu spacetime with the metric (\ref{PG-c}) with $M>0$ and $a(\tau)=a_0\tau^\alpha$ ($\alpha>0$).}
\begin{tabular}{|c|c|c|c|c|}
\hline
& $0<\alpha\le 1/2$ & $\alpha>1/2$ \\ \hline\hline
NEC & $0<\tau\le (\alpha+1)\sqrt{r^3/(2M)}$ & $0<\tau\le (\alpha+1)\sqrt{r^3/(2M)}$ \\ \hline
WEC & $0<\tau\le \alpha\sqrt{r^3/(2M)}$ & $0<\tau\le \alpha\sqrt{r^3/(2M)}$ \\ \hline
DEC & nowhere & $0<\tau\le [2(2\alpha-1)/7]\sqrt{r^3/(2M)}$ \\ \hline
SEC & $0<\tau\le (\alpha+1)\sqrt{r^3/(2M)}$ & $0<\tau\le (\alpha+1)\sqrt{r^3/(2M)}$ \\ \hline
\hline
\end{tabular}

\end{center}
\end{table}

In fact, the energy-momentum tensor (\ref{T-fluid-L}) {can be interpreted as} a combination of a homogeneous perfect fluid and an inhomogeneous anisotropic fluid such that $T_{\mu\nu}=T_{\mu\nu}^{\rm A}+T_{\mu\nu}^{\rm B}$, where
\begin{align}
&{T}_{\mu\nu}^{\rm A}=(\rho_F+p_F)u_\mu u_\nu+p_Fg_{\mu\nu}, \label{T-fluidA}\\
&{T}_{\mu\nu}^{\rm B}=({\bar \mu}+{\bar p}_{\rm t})u_\mu u_\nu+({\bar p}_{\rm r}-{\bar p}_{\rm t})s_\mu s_\nu +{\bar p}_{\rm t}g_{\mu\nu},\label{T-fluidB}\\
&8\pi \rho_F=3\frac{{\dot a}^2}{a^4}, \qquad 8\pi p_F=-2\frac{{\ddot a}}{a^3}+\frac{{\dot a}^2}{a^4},\\
&8\pi {\bar \mu}=-\frac{3{\dot a}}{a^3r}\sqrt{\frac{2M}{r}},\quad 8\pi {\bar p}_{\rm r}=\frac{4{\dot a}}{a^3r}\sqrt{\frac{2M}{r}},\quad 8\pi {\bar p}_{\rm t}=\frac{{\dot a}}{a^3r}\sqrt{\frac{2M}{r}}
\end{align}
with $u^\mu$ and $s^\mu$ given by Eq.~(\ref{us-Culetu}).
The inhomogeneous anisotropic fluid (\ref{T-fluidB}) satisfies equations of state ${\bar p}_{\rm r}=-4{\bar \mu}/3$ and ${\bar p}_{\rm t}=-{\bar \mu}/3$ and violates the NEC if ${\dot a}$ is non-zero by Eqs.~(\ref{NEC-I})--(\ref{SEC-I}).

Up to now, we have assumed the form of the conformal factor such as $a(\tau)=a_0\tau^\alpha$ ($\alpha>0$).
Even with a more general form of $a(\tau)$, it is shown under weak conditions that all the standard energy conditions are violated near the singularity $r=0$.
\begin{Prop}
\label{Prop:EC-Culetu}
Suppose ${\dot a}> 0$ and $M>0$ in the Culetu spacetime (\ref{PG-c}). Then, $r=0$ is a curvature singularity, around which the corresponding energy-momentum tensor in general relativity violates all the standard energy conditions.
\end{Prop}
{\it Proof:}
The first expression of the Ricci scalar (\ref{Ricci-s-PG}) shows that $r=0$ is a curvature singularity if ${\dot a}\ne 0$ and $M>0$ are satisfied.
Equations~(\ref{rho-Culetu})--(\ref{pt-Culetu}) give
\begin{align}
&\lim_{r\to 0}8\pi (\rho+p_{\rm t})\simeq -\frac{2{\dot a}}{a^3r}\sqrt{\frac{2M}{r}}\to -\infty,
\end{align}
so that all the standard energy conditions are violated near $r=0$ by Eqs.~(\ref{NEC-I})--(\ref{SEC-I}).
\qed

\subsection{Properties as a cosmological black hole}

Now we study a cosmological black hole described by the Culetu spacetime (\ref{PG-c}) with $a(\tau)=a_0\tau^\alpha~(\alpha>0)$ in the domain $\tau>0$ and $r>0$.

\subsubsection{Trapping horizon}

Here we identify the locations of trapping horizons and their types.
Consider a future-directed outgoing radial null vector $k^\mu$ and a future-directed ingoing radial null vector $l^\mu$ given by 
\begin{align}
k^\mu\frac{\partial}{\partial x^\mu}=&\frac{1}{\sqrt{2}a}\biggl\{\frac{\partial}{\partial \tau}+\biggl(1-\sqrt{\frac{2M}{r}}\biggl)\frac{\partial}{\partial r}\biggl\},\\
l^\mu\frac{\partial}{\partial x^\mu}=&\frac{1}{\sqrt{2}a}\biggl\{\frac{\partial}{\partial \tau}-\biggl(1+\sqrt{\frac{2M}{r}}\biggl)\frac{\partial}{\partial r}\biggl\},
\end{align}
which satisfy $k_\mu k^\mu=l_\mu l^\mu=0$ and $k_\mu l^\mu=-1$.
With the areal radius $R=ar$, the expansions (\ref{exp+}) and (\ref{exp-}) are computed to give
\begin{align}
\theta_+=&\frac{\sqrt{2}}{a}\biggl\{\frac{\alpha}{\tau}+\frac{1}{r}\biggl(1-\sqrt{\frac{2M}{r}}\biggl)\biggl\},\\
\theta_-=&\frac{\sqrt{2}}{a}\biggl\{\frac{\alpha}{\tau}-\frac{1}{r}\biggl(1+\sqrt{\frac{2M}{r}}\biggl)\biggl\},
\end{align}
which show that trapping horizons associated with $k^\mu$ and $l^\mu$ are respectively given by $\tau=\tau_+(r)$ and $\tau=\tau_-(r)$, where
\begin{align}
\tau_+(r):=-\alpha r\biggl(1-\sqrt{\frac{2M}{r}}\biggl)^{-1},\qquad \tau_-(r):=\alpha r\biggl(1+\sqrt{\frac{2M}{r}}\biggl)^{-1}.\label{tau-pm}
\end{align}
Note that $\tau_+(r)>0$ holds only in the domain $0<r<2M$ and then both $\tau_+(r)$ and $\tau_-(r)$ are monotonically increasing functions with $\tau_+(0)=\tau_-(0)=0$.

The line element along $\tau=\tau_+(r)$ on $(M^2,g_{AB})$ is given by 
\begin{align}
\D s^2|_{\tau=\tau_+}=&\frac{a(\tau_+)^2}{4(X-1)^{3}}\left\{2(1+\alpha)(X-1) +\alpha X\right\} \nonumber \\
&\times\left\{2(X-1)^{2}+2\alpha(X^2-1)+\alpha X(X+1)\right\}\D r^2,
\end{align}
where $X:=\sqrt{2M/r}$.
Since $0<r<2M$ gives $X>1$, $\D s^2|_{\tau=\tau_+}>0$ is satisfied and therefore $\tau=\tau_+(r)$ is spacelike.
On the other hand, the line element along $\tau=\tau_-(r)$ on $(M^2,g_{AB})$ is given by 
\begin{align}
\D s^2|_{\tau=\tau_-}=&\frac{a(\tau_-)^2}{4(X+1)^{3}}\left\{2(1+\alpha)(X+1) +\alpha X\right\}F(X)\D r^{2},
\end{align}
where
\begin{align}
F(X):= (2 + 3\alpha)X^2 + (4 - \alpha)X + 2 - 2\alpha.
\end{align}
The sign of $\D s^2|_{\tau=\tau_-}$ is determined by $F(X)$, which admits a positive real root for $\alpha>1$ and there is no positive real root for $0<\alpha\le 1$.
Thus, for $0<\alpha\le 1$, $\D s^2|_{\tau=\tau_-}>0$ is satisfied, so that $\tau=\tau_-(r)$ is spacelike.
For $\alpha>1$, $\tau=\tau_-(r)$ is timelike, null, and spacelike in the domains $r>r_{\rm d(C)}$, $r=r_{\rm d(C)}$, and $0<r<r_{\rm d(C)}$, respectively, where 
\begin{align}
r_{\rm d(C)}:=2M\biggl(\frac{2(2+3\alpha)}{\alpha-4+\sqrt{\alpha(25\alpha-16)}}\biggl)^2(>2M).\label{def-rdC}
\end{align}
The results are summarized in Table~\ref{table:signature-TH-Culetu}.
\begin{table}[htb]
\begin{center}
\caption{\label{table:signature-TH-Culetu} Signature of the trapping horizon $\tau=\tau_-(r)$ in the Culetu spacetime.}
\begin{tabular}{|c|c|c|c|c|}
\hline
& timelike & null & spacelike \\ \hline\hline
$0<\alpha\le 1$ & n.a. & n.a. & any $r$ \\ \hline
$\alpha>1$ & $r>r_{\rm d(C)}$ & $r=r_{\rm d(C)}$ & $0<r<r_{\rm d(C)}$ \\ 
\hline
\end{tabular}
\end{center}
\end{table}

\begin{figure}[htbp]
\begin{center}
\includegraphics[width=0.7\linewidth]{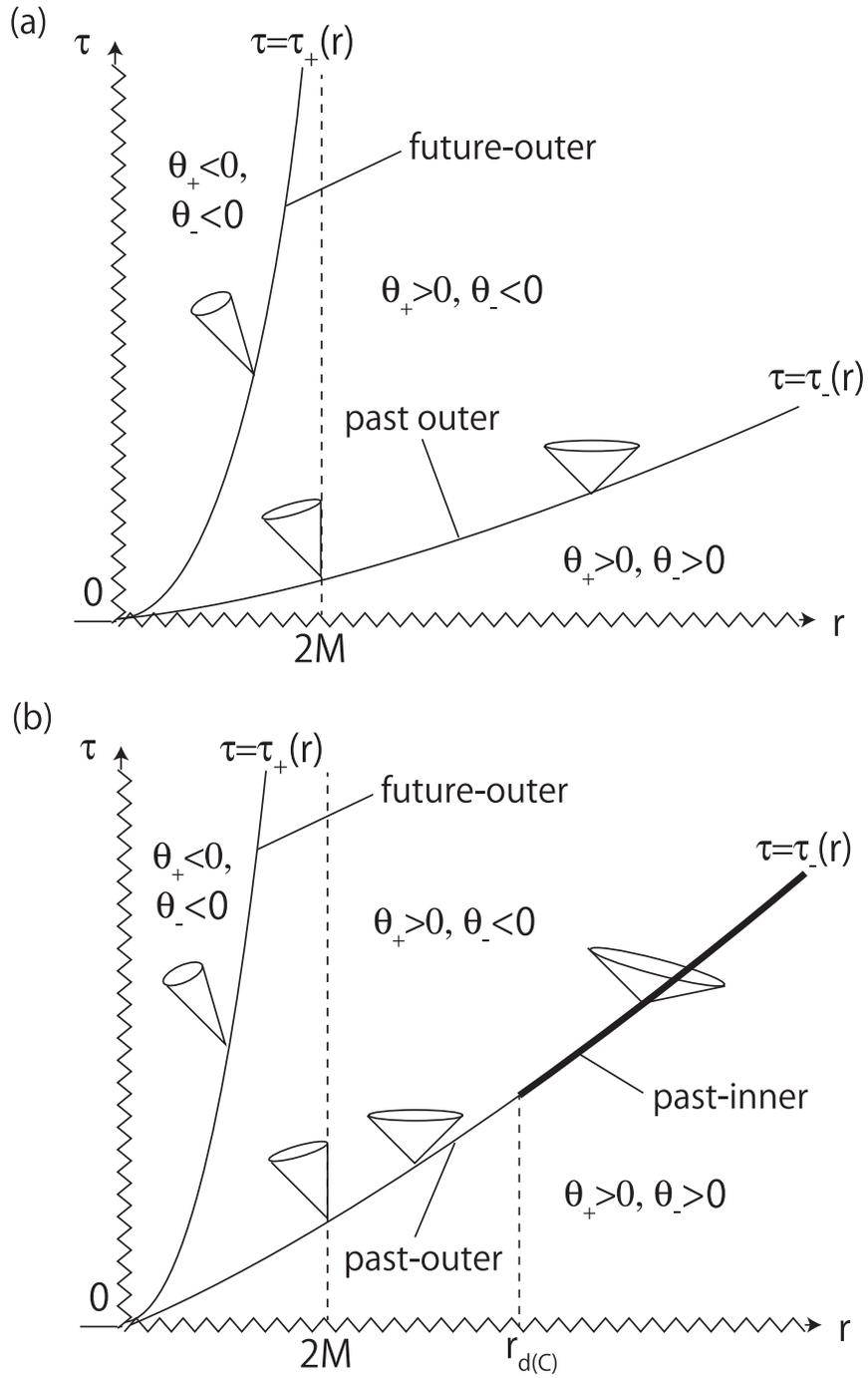}
\caption{\label{Fig:Culetu-r-tau-total} $(\tau,r)$ planes for the the Culetu spacetime with $a(\tau)=a_0\tau^{\alpha}$ for (a) $0<\alpha\le 1$ and (b) $\alpha>1$. 
Several future light-cones are put to clarify the signature of the trapping horizons $\tau=\tau_\pm(r)$. }
\end{center}
\end{figure}

In the $(\tau,r)$ plane shown in Fig.~\ref{Fig:Culetu-r-tau-total}, one recognizes that ${\cal L}_-\theta_+<0$ is satisfied at $\tau=\tau_+(r)$ defined by $\theta_+=0$ and therefore $\tau=\tau_+(r)$ is a future outer trapping horizon for any $\alpha(>0)$.
On the other hand, at $\tau=\tau_-(r)$ defined by $\theta_-=0$, ${\cal L}_+\theta_-<(>)0$ is satisfied when $\tau=\tau_-(r)$ is spacelike (timelike) and then $\tau=\tau_-(r)$ is a past outer (past inner) trapping horizon.
Therefore, the trapping horizon $\tau=\tau_-(r)$ is past outer for $0<\alpha\le 1$.
For $\alpha>1$, $\tau=\tau_-(r)$ is past outer, past degenerate, and past inner in the domain $0<r<r_{\rm d(C)}$, $r=r_{\rm d(C)}$, and $r>r_{\rm d(C)}$, respectively.

\begin{figure}[htbp]
\begin{center}
\includegraphics[width=0.5\linewidth]{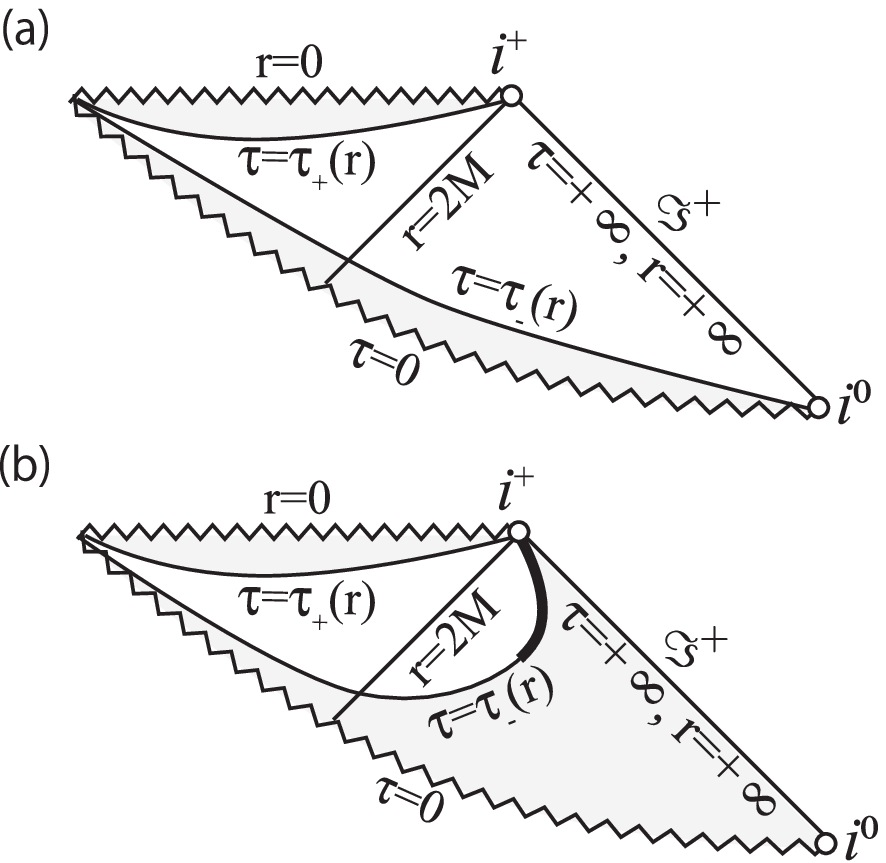}
\caption{\label{Fig:Culetu-TH} Penrose diagrams of the Culetu black-hole spacetime with $a(\tau)=a_0\tau^{\alpha}$ for (a) $0<\alpha\le 1$ and 
(b) $\alpha>1$. 
Shaded regions are trapped regions. A thick timelike portion of $\tau=\tau_-(r)$ is a past inner trapping horizon. }
\end{center}
\end{figure}

Collecting all the information obtained, one can draw the Penrose diagrams of the Culetu black hole as in Fig.~\ref{Fig:Culetu-TH}.
We confirm the asymptotic behaviors of the trapping horizons $\tau=\tau_\pm(r)$ as follows.
Using Eqs.~(\ref{uv-def}), (\ref{r*-def}), (\ref{UV-def}), (\ref{UV-r}), and (\ref{def:tau-PG}), one can rewrite the compactified Kruskal-Szekeres coordinates (\ref{barUV-def}) in terms of $(\tau,r)$ as
\begin{align}
{\bar U}=&\arctan\biggl[\biggl(1-\sqrt{\frac{r}{2M}}\biggl)\exp\biggl(\frac{r-\tau}{4M}+\sqrt{\frac{r}{2M}}\biggl)\biggl],\label{barU-Culetu}\\
{\bar V}=&\arctan\biggl[\biggl(1+\sqrt{\frac{r}{2M}}\biggl)\exp\biggl(\frac{r+\tau}{4M}-\sqrt{\frac{r}{2M}}\biggl)\biggl].\label{barV-Culetu}
\end{align}
Substituting $\tau=\tau_+(r)$ defined by Eq.~(\ref{tau-pm}), we obtain
\begin{align}
{\bar U}=&\arctan\biggl[\biggl(1-\sqrt{\frac{r}{2M}}\biggl)\exp\biggl(\frac{r}{4M}+\frac{\alpha r}{2M}\biggl(1-\sqrt{\frac{2M}{r}}\biggl)^{-1}+\sqrt{\frac{r}{2M}}\biggl)\biggl],\\
{\bar V}=&\arctan\biggl[\biggl(1+\sqrt{\frac{r}{2M}}\biggl)\exp\biggl(\frac{r}{4M}-\frac{\alpha r}{2M}\biggl(1-\sqrt{\frac{2M}{r}}\biggl)^{-1}-\sqrt{\frac{r}{2M}}\biggl)\biggl].
\end{align}
The above expressions show $({\bar U},{\bar V})\to (0,\pi/2)$ as $r\to 2M^-$. (See Fig.~\ref{Culetu}.)
On the other hand, substituting $\tau=\tau_-(r)$ defined by Eq.~(\ref{tau-pm}) into Eqs.~(\ref{barU-Culetu}) and (\ref{barV-Culetu}), we obtain
\begin{align}
{\bar U}=&\arctan\biggl[\biggl(1-\sqrt{\frac{r}{2M}}\biggl)\exp\biggl(\frac{r}{4M}-\frac{\alpha r}{4M}\biggl(1+\sqrt{\frac{2M}{r}}\biggl)^{-1}+\sqrt{\frac{r}{2M}}\biggl)\biggl],\\
{\bar V}=&\arctan\biggl[\biggl(1+\sqrt{\frac{r}{2M}}\biggl)\exp\biggl(\frac{r}{4M}+\frac{\alpha r}{4M}\biggl(1+\sqrt{\frac{2M}{r}}\biggl)^{-1}-\sqrt{\frac{r}{2M}}\biggl)\biggl].
\end{align}
The asymptotic behaviors as $r \to \infty$ are $({\bar U},{\bar V}) \to(-\pi/2,\pi/2)$ for $0 < \alpha \leq 1$ and $({\bar U},{\bar V}) \to(0,\pi/2)$ for $\alpha>1$.
Hence, $\tau=\tau_-(r)$ approaches the spacelike infinity $i^0$ in Fig.~\ref{Culetu} for $0 < \alpha \leq 1$ and the future timelike infinity $i^+$ for $\alpha>1$.

\subsubsection{Misner-Sharp mass}

The Misner-Sharp mass (\ref{MS-mass}) for the Culetu spacetime with the metric (\ref{PG-c}) is computed to give
\begin{align}
m_{\rm MS}=\frac{ar}{2}\biggl(\sqrt{\frac{2M}{r}}-\frac{{\dot a}}{a}r\biggl)^2=\frac{a_0\tau^{\alpha-2} r}{2}\biggl(\tau\sqrt{\frac{2M}{r}}-\alpha r\biggl)^2,\label{MSmass-PG}
\end{align}
which is non-negative everywhere.
At the central singularity $r=0$ and on the event horizon ($r=2M$), the Misner-Sharp mass is given by $m_{\rm MS}(\tau,0)=a(\tau)M$ and 
\begin{align}
m_{\rm MS}(\tau,2M)=Ma_0\tau^{\alpha-2}(\tau-2\alpha M)^2,
\end{align}
respectively.

Differentiating Eq.~(\ref{MSmass-PG}) with respect to $r$, we obtain
\begin{align}
\partial_r{m_{\rm MS}}=\frac{3{\dot a}r}{2}\biggl(\frac{{\dot a}}{a}r-\sqrt{\frac{2M}{r}}\biggl)=\frac{3\alpha a_0\tau^{\alpha-1}r}{2}\biggl(\frac{\alpha r}{\tau}-\sqrt{\frac{2M}{r}}\biggl),
\end{align}
which shows $\partial_rm_{\rm MS}<0$ in the region of 
\begin{align}
\tau>\alpha r\sqrt{\frac{r}{2M}}.
\end{align}
Figures~\ref{Fig:Culetu-r-tau-total} and \ref{Fig:Culetu-TH} show that the event horizon $r=2M$ is in an untrapped region for $\tau>\tau_-(2M)(=\alpha M)$.
Since the contraposition of Proposition~\ref{th:monotonicity} asserts that the DEC is violated in an untrapped region with $\partial_r{m_{\rm MS}}<0$, the DEC is violated on the event horizon $r=2M$ for $\tau>2\alpha M$.
This is consistent with Table~\ref{table:EC-Culetu} showing that the DEC is violated everywhere for $0<\alpha\le 1/2$ and on the event horizon in the late time $\tau> 4(2\alpha-1)M/7$ for $\alpha>1/2$ because $2\alpha M>4(2\alpha-1)M/7$ is satisfied.

\subsubsection{Temperature of a cosmological black hole}

Since the conformal Killing vector $\xi^\mu=(1,0,0,0)$ in the Culetu spacetime with the metric (\ref{PG-c}) satisfies ${\cal L}_\xi g_{\mu\nu}=2({\dot a}/a) g_{\mu\nu}$ and $\xi_\mu \xi^\mu=-a^2(1-2M/r)$, the event horizon $r=2M$ is a conformal Killing horizon.
Its temperature (\ref{T-CK}) is computed to give
\begin{align}
T_{\rm JKSD}=\frac{1}{8\pi M},
\end{align}
which is the same as the temperature of the Schwarzschild black hole. 

Next, we derive the temperature of the future outer trapping horizon $\tau=\tau_+(r)$.
In the Culetu spacetime (\ref{PG-c}), the Kodama vector (\ref{kodamavector}) is given by 
\begin{align}
K^\mu\frac{\partial}{\partial x^\mu}=\frac{1}{a}\frac{\partial}{\partial \tau}-\frac{r{\dot a}}{a^2}\frac{\partial}{\partial r}.
\end{align}
The temperature (\ref{T-TH}) of $\tau=\tau_+(r)$ defined by Eq.~(\ref{tau-pm}) is computed to give
\begin{align}
T_{\rm TH}=&\frac{1}{4\pi a_0\alpha r^2}\sqrt{\frac{r}{2M}}\biggl\{\frac{1}{\alpha r}\biggl(\sqrt{\frac{2M}{r}}-1\biggl)\biggl\}^\alpha \nonumber \\
&\times\biggl[\left\{(3\alpha+2) M+(1-\alpha)r\right\}\sqrt{\frac{2M}{r}}+M(\alpha-4)\biggl] \nonumber \\
=&\frac{1}{8\pi M}\times\frac{j(w)}{a_0(2M)^{\alpha}},
\end{align}
where $j(w)$ is a dimensionless function of $w:=\sqrt{r/(2M)}$ defined by 
\begin{align}
j(w)=&\frac{(1-w)^\alpha[2(1-\alpha)w^2+(\alpha-4)w+(3\alpha+2)]}{2\alpha^{1+\alpha}w^{4+3\alpha}}.
\end{align}
The function $j(w)$ can be shown by cases to be positive in the domain $0<w<1$ (corresponding to $0<r<2M$) for $\alpha>0$.
Hence, $T_{\rm TH}$ is a positive function of $r$ in the domain $0<r<2M$ and it diverges as $r\to 0$ ($\tau\to 0$) and converges to zero as $r\to 2M$ ($\tau\to\infty$).

In this section, we have fully investigated the Culetu spacetime and the corresponding matter field.
Actually, the Culetu spacetime can be slightly modified to be non-conformally Schwarzschild as shown in Appendix~\ref{app:mod-Culetu}, which is also a candidate of a cosmological black-hole spacetime.

\section{Summary and discussions}
\label{sec:summary}

\subsection{Summary}
In the present paper, we have fully investigated various conformally Schwarzschild spacetimes which are asymptotically flat FLRW universe filled by a perfect fluid obeying a linear equation state $p=w\rho$ with $w>-1/3$.
Among them, as shown in Sec.~\ref{sec:fail}, the Thakurta spacetime (\ref{Thakurta1}) constructed with the standard Schwarzschild coordinates and the McClure-Dyer spacetime (\ref{MD}) constructed with the isotropic coordinates are identical.
Therefore, according to the results in~\cite{Mello:2016irl,Harada:2021xze}, these spacetimes do not describe a cosmological black hole.

In Sec.~\ref{sec:SD}, we have clarified that the region with $\eta>0$ and $r>0$ of the Sultana-Dyer class of conformally Schwarzschild spacetimes (\ref{metricSD}) constructed with the Kerr-Schild coordinates and $a(\eta) = a_{0}|\eta|^{\alpha}~(\alpha>0)$ describe a cosmological black hole, where $\eta=0$ and $r=0$ are curvature singularities.
We have shown that the corresponding matter field in this cosmological black-hole spacetime can be interpreted as a combination of a homogeneous perfect fluid and an inhomogeneous null fluid.
Different from the interpretation by Sultana and Dyer~\cite{Sultana:2005tp} as a combination of a perfect fluid and a null dust, this novel interpretation of matter is valid in the whole spacetime.
While the homogeneous perfect fluid is identical to the one in the background FLRW universe, the inhomogeneous type-II null fluid violates the NEC near the black-hole singularity at $r=0$ as shown in Table~\ref{table:EC-SD}.
As summarized in Table~\ref{table:EC-SD-total}, the total energy-momentum tensor violate all the standard energy conditions in the region with a finite $r$ for sufficiently large $\eta$.
We have also shown that the domain $\eta<0$ also describes a cosmological black hole if the coordinate boundary given by $(\eta,r)\to (-\infty,2M)$ is regular.

In Sec.~\ref{sec:Culetu}, we have clarified the global structure of the Culetu spacetime with the metric (\ref{PG-c}) constructed with the Painlev\'{e}-Gullstrand coordinates and $a(\tau) = a_{0}|\tau|^{\alpha}~(\alpha>0)$ and shown that the region with $\tau>0$ and $r>0$ describes a cosmological black hole, where $\tau=0$ and $r=0$ are curvature singularities.
As Culetu pointed out in~\cite{Culetu:2012ih}, the corresponding matter field in this cosmological black-hole spacetime can be interpreted as a {single} anisotropic fluid and can also be interpreted as a combination of a homogeneous (cosmological) perfect fluid and an inhomogeneous anisotropic fluid.
The latter inhomogeneous anisotropic fluid obeys linear equations of state and violates all the standard energy conditions everywhere.
Similarly to the Sultana-Dyer class of cosmological black holes, the total energy-momentum tensor violate all the standard energy conditions in the region with a finite $r$ for sufficiently large $\tau$ as shown in Table~\ref{table:EC-Culetu}.
We have also shown that the domain $\tau<0$ in the Culetu spacetime describes a cosmological black hole if the coordinate boundary given by $(\tau,r)\to (-\infty,2M)$ is regular.

\subsection{Discussions}
In the present paper, it has been clarified that the Sultana-Dyer class of spacetimes and the Culetu spacetime describe a cosmological black hole in the decelerating flat FLRW universe.
However, they share a crucial property that the energy conditions are initially satisfied but later violated as solutions of the Einstein equations. 
In concluding this paper, we will show that the NEC is later violated on the event horizon, and consequently neither of these classes of spacetimes describe the evolution of a primordial black hole in general relativity after it gets smaller than the Hubble horizon.
Since the qualitative properties of the Sultana-Dyer and Culetu cosmological black holes are similar, we will handle them together below.
The conformal time $t$ in the following argument stands for $\eta$ and $\tau$ in the Sultana-Dyer metric (\ref{metricSD}) and Culetu metric (\ref{PG-c}), respectively.

Here we express the asymptotic background FLRW universe in terms of the cosmological time ${\bar t}$ as
\begin{eqnarray}
\D s^2=-\D{\bar t}^2+{\bar a}({\bar t})^2(\D r^2+r^2\D\Omega^2),
\end{eqnarray}
which is obtained from the metric (\ref{flat-FLRW}) with the conformal time $t$ by a coordinate transformation $t=t({\bar t})$ defined by $a(t)\D t=\D {\bar t}$ and a redefinition of the scale factor as ${\bar a}({\bar t}):=a(t({\bar t}))$.
If the background FLRW universe is filled with a perfect fluid obeying an equation of state $p=w\rho$, we have ${\bar a}({\bar t})=b_{0}{\bar t}^{\beta}$, where $b_0$ is a positive constant and $\beta$ is given by $\beta=2/[3(1+w)]$.
In the metric (\ref{flat-FLRW}) with the conformal time $t$, we have $a(t)=a_0t^\alpha$ with $\alpha=2/(3w+1)$, and thus $\beta=\alpha/(1+\alpha)$ holds.

Note that $w=0$ and $w=1/3$ correspond to a dust fluid and a radiation fluid, respectively, and the DEC for the background universe requires $-1\le w\le 1$.
Our assumption $w>-1/3$ in this paper corresponds to $\alpha>0$ and $0<\beta<1$.
In the following, we assume $-1/3<w\le 1$, or equivalently $1/3\le \beta<1$, under which the background FLRW universe is decelerating and the DEC is satisfied there.
Then, the relation between ${\bar t}$ and $t$ is
\begin{equation}
{\bar t}=[(1-\beta)b_{0}t]^{1/(1-\beta)}~~\leftrightarrow~~b_{0} t=\frac{1}{1-\beta}{\bar t}^{1-\beta},\label{eq:t_eta}
\end{equation}
which is obtained by integrating $\D t=\D {\bar t}/{\bar a}({\bar t})$.

The location of the event horizon is given by $r=2M$ both for the Sultana-Dyer and Culetu cosmological black holes.
For the Sultana-Dyer black hole, the NEC is violated (and hence all the standard energy conditions are violated) on and outside the event horizon in the period of $\eta>28(\alpha+1)M/9(=\eta_{\rm N}(2M))$, where $\eta_{\rm N}(r)$ is defined by Eq.~(\ref{NEC-SD}).
For the Culetu black hole, as shown in Table~\ref{table:EC-Culetu}, this period of the NEC violation is given by $\tau>2(\alpha+1)M$.
By Eq.~(\ref{eq:t_eta}), we express these two inequalities in a unified manner in terms of the cosmological time as ${\bar t}>{\bar t}_{\rm V}$, where
\begin{equation}
{\bar t}_{\rm V}= (2qMb_{0})^{1/(1-\beta)} \label{eq:t_V}
\end{equation}
with $q=14/9$ for Sultana-Dyer and $q=1$ for Culetu.

The areal radius of the event horizon is given by $R({\bar t},2M)=2M{\bar a}({\bar t})$.
Since the expansion of the background universe is decelerated, the event horizon initially larger but subsequently becomes smaller
than the Hubble horizon $H^{-1}(={\bar a}/{\dot {\bar a}}={\bar t}/\beta)$.
In cosmology, the formation time of a primordial black hole is usually identified in order estimation with the time when the event horizon ``enters'' the Hubble horizon~\cite{Carr:1975qj}.
Thus, the formation time ${\bar t}={\bar t}_{\rm F}$ is determined by $H^{-1}=2M{\bar a}({\bar t})$, which is solved to give
\begin{equation}
{\bar t}_{\rm F}=(2\beta Mb_0)^{1/(1-\beta)}. \label{eq:t_F}
\end{equation}

From Eqs.~(\ref{eq:t_V}) and (\ref{eq:t_F}), we conclude ${\bar t}_{\rm F}\simeq {\bar t}_{\rm V}$ for $1/3\le \beta<1$ (or equivalently $-1/3<w\le 1$), i.e., all the standard energy conditions are violated on the event horizon as soon as the primordial black hole forms.
In other words, if the NEC is imposed on and outside the event horizon, neither the Sultana-Dyer nor Culetu metric describes the evolution of a primordial black hole after the horizon entry.
Although ${\bar t}_{\rm F}\ll {\bar t}_{\rm V}$ is realized for $\beta\ll 1$, it corresponds to $w\gg 1$ and then the DEC is violated in the background FLRW universe.
To summarize, the results obtained in this paper firmly undermine the validity of the Sultana-Dyer and Culetu metrics as a model of primordial black holes.

\subsection*{Acknowledgements}
TS and TH are very grateful to Y.~Hatsuda, M.~Kimura T.~Kobayashi, Y.~Koga, H.~Motohashi, Y.~Nakayama and C.-M.~Yoo for fruitful 
discussions and helpful comments.
Special thanks are due to J. M. M. Senovilla for reading the published version carefully and letting them know some points to be corrected.
This work was supported in part by Japan Society for the Promotion of Science (JSPS) KAKENHI Grant No. JP19K03876 (TH), No. JP19H01895 (TH), and No. JP20H05853 (TH).

\appendix

\section{Other candidates: Non-conformally Schwarzschild spacetimes}
\label{app:others}

The McVittie solution~\cite{McVittie:1933zz} is a spherically symmetric solution with a perfect fluid which is asymptotic to the FLRW spacetime with positive, zero, or negative spatial curvature.
In particular, the asymptotically flat FLRW McVittie solution is given by
\begin{align}
\label{McVittie}
&\D s^2=-\frac{(2a(t)\sigma-M)^2}{(2a(t)\sigma+M)^2}\D t^2+a(t)^2\biggl(1+\frac{M}{2a(t)\sigma}\biggl)^4\left(\D \sigma^2+\sigma^2\D \Omega^2\right),\\
&T_{\mu\nu}=(p+\rho)u_\mu u_\nu+pg_{\mu\nu}
\end{align}
with constant $M$ and arbitrary function $a(t)$, where $u^\mu$, $\rho$, and $p$ are given by
\begin{align}
&u^\mu\frac{\partial}{\partial x^\mu}=\biggl|\frac{2a\sigma+M}{2a\sigma-M}\biggl|\frac{\partial}{\partial t},\qquad 8\pi \rho=3\frac{{\dot a}^2}{a^2},\label{em00}\\
&8\pi p=-2\frac{\ddot a}{a}\frac{(2a\sigma+M)}{(2a\sigma-M)}-\frac{{\dot a}^2}{a^2}\frac{(2a\sigma-5M)}{(2a\sigma-M)}.\label{em11}
\end{align}
The spacetime (\ref{McVittie}) is asymptotic to the flat FLRW spacetime as $\sigma\to \infty$.
It reduces to the flat FLRW spacetime and the Schwarzschild spacetime in the isotropic coordinates (\ref{Sch-iso}) for $M=0$ and $a(t)\equiv 1$, respectively.

Since the McVittie spacetime (\ref{McVittie}) is not conformally Schwarzschild, it is not an easy task to clarify the global structure of the spacetime.
After a huge effort by Nolan in his series of papers~\cite{Nolan:1998xs,Nolan:1999kk,Nolan:1999wf}, it has been finally shown that the coordinate system (\ref{McVittie}) does not cover a maximally extended spacetime if the scale factor $a(t)$ obeys $a(t)\propto \exp(H_0 t)$ as $t\to \infty$ with a positive constant $H_0$, where the maximally extended spacetime describes a cosmological black hole~\cite{Kaloper:2010ec,Lake:2011ni}.
In this appendix, we present another two non-conformally Schwarzschild spacetimes as candidates of a cosmological black-hole spacetime.

\subsection{Conformally Husain spacetime}
\label{app:Husain}

In~\cite{Husain:1995bf}, Husain showed that the following metric
\begin{align}
\label{husain}
\D s^2=-\D\eta^2+\D r^2+r^2\D\Omega^2+\frac{2M(\eta,r)}{r}(\D \eta+\D r)^2
\end{align}
solves the Einstein equations with a type-II null fluid $T^{\mu\nu}=\Omega l^\mu l^\nu+(\mu+P)(l^\mu n^\nu+n^\mu l^\nu )+Pg^{\mu\nu}$, where $l_\mu l^\mu =n_\mu n^\mu=0$ and $l_\mu n^\mu=-1$ are satisfied.
By a coordinate transformation $v:=\eta+r$, the Husain spacetime (\ref{husain}) is written as
\begin{align}
\label{husain2}
\D s^2=-\biggl(1-\frac{2{\bar M}(v,r)}{r}\biggl)\D v^2+2\D v\D r+r^2\D\Omega^2,
\end{align}
where ${\bar M}(v,r):=M(\eta(v,r),r)$.
Clearly the spacetimes with the metric~(\ref{husain}) or (\ref{husain2}) are generalizations of the Schwarzschild solution in the Kerr-Schild coordinates~(\ref{Sch-KS}) and the ingoing Eddington-Finkelstein coordinates (\ref{Sc3}), respectively.
The Husain solution (\ref{husain2}) is also dubbed the generalized Vaidya solution as it reduces to the Vaidya solution with a null dust $T^{\mu\nu}=\Omega l^\mu l^\nu$ with ${\bar M}(v,r)={\bar M}(v)$.

Now we consider the following conformally Husain spacetime as a generalization of the Sultana-Dyer spacetime (\ref{metricSD}):
\begin{eqnarray}
\D s^2&=&a(\eta)^2\biggl[-\D \eta^2+\D r^2+r^2\D\Omega^2+\frac{2M(\eta,r)}{r}(\D \eta+\D r)^2\biggl].\label{c-husain}
\end{eqnarray}
In fact, this metric solves the Einstein equations with the following energy-momentum tensor that is a combination of a perfect fluid and a type II null fluid:
\begin{align}
T^{\mu\nu}=&T_{\mathcal{A}}^{\mu\nu}+T_{\mathcal{B}}^{\mu\nu},\label{emt-husain2}\\
T_{\mathcal{A}}^{\mu\nu}:=&\rho_{F} {\tilde u}^\mu {\tilde u}^\nu+p_{F}(g^{\mu\nu}+{\tilde u}^\mu {\tilde u}^\nu),\\
T_{\mathcal{B}}^{\mu\nu}=&\Omega l^\mu l^\nu+(\mu+P)(l^\mu n^\nu+l^\nu n^\mu)+Pg^{\mu\nu},
\end{align}
where
\begin{align}
&{\tilde u}_{\mu}\D x^\mu=-a\biggl[\biggl(1-\frac{M(\eta,r)}{r}\biggl)\D \eta-\frac{M(\eta,r)}{r}\D r\biggl],\\
&8\pi\rho_{F}=\frac{3{\dot a}^2}{a^4},\qquad 8\pi p_{F}=\frac{{\dot a}^2}{a^4}-\frac{2{\ddot a}}{a^3}.
\end{align}
and
\begin{align}
&l_\mu\D x^\mu=-\frac{1}{\sqrt{2}}a(\D \eta+\D r),\\
&n_{\mu}\D x^\mu =-\frac{1}{\sqrt{2}}a\biggl[\biggl(1-\frac{2M(\eta,r)}{r}\biggl)\D \eta-\biggl(1+\frac{2M(\eta,r)}{r}\biggl)\D r\biggl],\\
&8\pi\mu=\frac{2M}{r^2}\biggl(r\frac{{\dot a}^2}{a^4}+r\frac{{\ddot a}}{a^3}-\frac{3{\dot a}}{a^3}\biggl)+\frac{2({\dot M}-M')}{r^2}\biggl(r\frac{{\dot a}}{a^3}-\frac{1}{a^2}\biggl),\\
&4\pi\Omega=\frac{2M}{r^2}\biggl(2(r+M)\frac{{\dot a}^2}{a^4}-(r+M)\frac{{\ddot a}}{a^3}+\frac{{\dot a}}{a^3}\biggl){+}\frac{2{\dot M}}{r^2a^2}{-}\frac{2{\dot a}M'}{ra^3},\\
&8\pi P=\frac{2M}{r}\biggl(\frac{{\dot a}^2}{a^4}-\frac{2{\ddot a}}{a^3}\biggl)-\frac{4{\dot a}({\dot M}-M')}{ra^3}-\frac{{\ddot M}+M''-2{\dot M}'}{ra^2}.\label{emt-husain2-last}
\end{align}
Here ${\tilde u}_\mu {\tilde u}^\mu=-1$, $l_\mu l^\mu=n_\mu n^\mu=0$, and $l_\mu n^\mu=-1$ are satisfied.
With $M(\eta,r)=M=$constant, Eqs.~(\ref{emt-husain2})--(\ref{emt-husain2-last}) reduce to the energy-momentum tensor (\ref{emt-SD2}) for the Sultana-Dyer spacetime.

\subsection{Modified Culetu spacetime}
\label{app:mod-Culetu}

As a non-conformally Schwarzschild modification of the Culetu spacetime in the Lema\^{\i}tre coordinates (\ref{Lemaitre-c}), we consider the following metric:
\begin{eqnarray}
\D s^2=-\D {\bar \tau}^2+a({\bar \tau})^2\biggl\{(2M)^{2/3}\biggl(\frac{\D \zeta^2}{\left[\frac32(\zeta-{\bar \tau})\right]^{2/3}}+\left[\frac32(\zeta-{\bar \tau})\right]^{4/3}\D \Omega^2\biggl)\biggl\}. \label{generalized-C}
\end{eqnarray}
Similar to the Culetu spacetime, ${\bar\tau}=\zeta$ is a curvature singularity.
Also, ${\bar \tau}$ is still timelike everywhere and the Einstein tensor $G^\mu_{~\nu}$ is diagonal such that
\begin{align}
G^{{\bar \tau}}_{~{\bar \tau}}=&-3\frac{{\dot a}^2}{a^2}+\frac{2{\dot a}}{a(\zeta-{\bar \tau})},\\
G^{\zeta}_{~\zeta}=&-2\frac{{\ddot a}}{a}-\frac{{\dot a}^2}{a^2}+\frac{4{\dot a}}{a(\zeta-{\bar \tau})},\\
G^{\theta}_{~\theta}=&G^{\phi}_{~\phi}=-2\frac{{\ddot a}}{a}-\frac{{\dot a}^2}{a^2}+\frac{{\dot a}}{a(\zeta-{\bar \tau})}.
\end{align}
Therefore, the corresponding energy-momentum tensor $T_{\mu\nu}$ may be interpreted as a combination of a homogeneous perfect fluid and an inhomogeneous anisotropic fluid such that $T_{\mu\nu}=T_{\mu\nu}^{\rm A}+T_{\mu\nu}^{\rm B}$, where
\begin{align}
&{T}_{\mu\nu}^{\rm A}=(\rho_F+p_F)u_\mu u_\nu+p_Fg_{\mu\nu},\qquad u^\mu\frac{\partial}{\partial x^\mu}=\frac{\partial}{\partial {\bar \tau}}, \label{T-fluidA2}\\
&8\pi \rho_F=3\frac{{\dot a}^2}{a^2}, \qquad 8\pi p_F=-2\frac{{\ddot a}}{a}-\frac{{\dot a}^2}{a^2}
\end{align}
and
\begin{align}
&{T}_{\mu\nu}^{\rm B}=(\mu+p_{\rm t})u_\mu u_\nu+(p_{\rm r}-p_{\rm t})s_\mu s_\nu +p_{\rm t}g_{\mu\nu},\label{T-fluidB2}\\
&u^\mu\frac{\partial}{\partial x^\mu}=\frac{\partial}{\partial {\bar \tau}},\qquad s^\mu\frac{\partial}{\partial x^\mu}=\frac{1}{a}\left[\frac{3}{4M}(\zeta-{\bar \tau})\right]^{1/3}\frac{\partial}{\partial \zeta},\\
&8\pi \mu=-\frac{2{\dot a}}{a(\zeta-{\bar \tau})},\quad 8\pi p_{\rm r}=\frac{4{\dot a}}{a(\zeta-{\bar \tau})},\quad 8\pi p_{\rm t}=\frac{{\dot a}}{a(\zeta-{\bar \tau})}.
\end{align}
The inhomogeneous anisotropic fluid (\ref{T-fluidB2}) satisfies equations of state $p_{\rm r}=-2\mu$ and $p_{\rm t}=-\mu/2$, which violates the NEC if ${\dot a}$ is non-zero by Eqs.~(\ref{NEC-I})--(\ref{SEC-I}).

\end{document}